\documentstyle[aps]{revtex}
\tighten
\begin{document}


\preprint{HUTP-01/A064,hep-th/0112205}

\title{Brane gravity, higher derivative terms and non-locality}
\author{Shinji Mukohyama}
\address{
Department of Physics, Harvard University\\
Cambridge, MA, 02138, USA
}
\date{\today}

\maketitle

\begin{abstract} 
 In brane world scenarios with a bulk scalar field between two branes it
 is known that $4$-dimensional Einstein gravity is restored at low
 energies on either brane. By using a gauge-invariant gravitational and
 scalar perturbation formalism we extend the theory of weak gravity in
 the brane world scenarios to higher energies, or shorter distances. We
 argue that weak gravity on either brane is indistinguishable from
 $4$-dimensional higher derivative gravity, provided that the inter-brane
 distance (radion) is stabilized, that the background bulk scalar field
 is changing near the branes and that the background bulk geometry near
 the branes is warped. This argument holds for a general conformal
 transformation to a frame in which matter on the branes is minimally
 coupled to the metric. In particular, Newton's constant and the
 coefficients of curvature-squared terms in the $4$-dimensional
 effective action are determined up to an ambiguity of adding a
 Gauss-Bonnet topological term. In other words, we provide the
 brane-world realization of the so called $R^2$-model without utilizing
 a quantum theory. We discuss the appearance of composite spin-$2$ and
 spin-$0$ fields in addition to the graviton on the brane and point out
 a possibility that the spin-$0$ field may play the role of an effective
 inflaton to drive brane-world inflation. Finally, we conjecture that
 the sequence of higher derivative terms is an infinite series and,
 thus, indicates non-locality in the brane world scenarios. 
\end{abstract}



\section{Introduction}


In the recent development of string/M theory~\cite{Polchinski}, branes
have been playing many important roles. The idea that our universe is a
brane in a higher dimensional spacetime has been attracting a great deal
of interest~\cite{ADD,AADD,RS1,RS2}. Although the idea of the brane
world had arisen at a phenomenological level already in
1983~\cite{earlier-works}, it is perhaps the discovery of the duality
between M-theory and $E_8\times E_8$ heterotic superstring theory by
Horava and Witten~\cite{Horava} that made it more attractive. It
actually gives the brane world idea a theoretical background: by
compactifying six dimensions in the $11$-dimensional theory, our
$4$-dimensional universe may be realized as a hypersurface in
$5$ dimensions at one of the fixed points of an $S_1/Z_2$
compactification. After compactification of $6$ dimensions by a
Calabi-Yau manifold, the $5$-dimensional effective theory can be
obtained, see e.g. \cite{Lukas}. 


Randall and Sundrum proposed two similar but distinct phenomenological
brane world scenarios~\cite{RS1,RS2}. In the scenario the
$5$-dimensional spacetime is compactified on $S_1/Z_2$ and all matter
fields are assumed to be confined on branes at fixed points of the
$S_1/Z_2$ so that the bulk, or the spacetime region between two fixed
points, is described by pure Einstein gravity with a negative
cosmological constant. In the second scenario the fifth dimension is
infinite but still the $Z_2$ symmetry is imposed. In both scenarios the
existence of the branes and the bulk cosmological constant makes the
bulk geometry curved, or warped. There are generalizations of their
scenarios with a scalar field between two branes~\cite{GW,DFGK}. In
these generalized warped brane-world scenarios, the scalar field was
introduced to stabilize a modulus called the radion, which represents a
separation between the two branes.


In the brane world scenarios with or without a scalar field in a warped
bulk geometry, weak gravity in a static background has been extensively 
investigated~\cite{Garriga-Tanaka,SSM,GKR,Tanaka-Montes,Kudoh-Tanaka,Mukohyama-Kofman}.
It is known that weak gravity in the scenario without a bulk scalar
between two branes is not Einstein but a Brans-Dicke theory at low
energies. On the other hand, in scenarios with a bulk scalar field
between two branes $4$-dimensional Einstein gravity is restored at low
energies. It is believed that the validity of Einstein gravity breaks
down at a certain energy scale which can be much lower than
$4$-dimensional Planck energy.


Hence, it seems natural to ask 'what does gravity in brane worlds
look like at high energies or at short distances?' In other words, 'how 
does the $4$-dimensional description break down?'


In this paper we investigate weak gravity in brane world scenarios with
a bulk scalar field between two branes at higher energies. For this
purpose we use the gauge-invariant perturbation formalism developed in
ref.~\cite{Mukohyama-Kofman}. In this formalism all quantities and
equations are Fourier transformed with respect to the $4$-dimensional
coordinates and classified into scalar, vector and tensor perturbations
so that the problem is reduced to a set of purely $1$-dimensional
problems. We also adopt expansion in a parameter $\mu\equiv
l^2\eta^{\mu\nu}k_{\mu}k_{\nu}$, where $l$ is a characteristic length
scale of the model and $k_{\mu}$ is the $4$-dimensional momentum (or the
Fourier parameter). In the lowest order in $\mu$, $4$-dimensional
Einstein gravity is restored on either brane~\cite{Mukohyama-Kofman}. In
the next order it is shown that gravity on either brane is
indistinguishable from a higher derivative gravity whose action includes
the Einstein term and curvature-squared terms. Equipped with the result
for this order, we conjecture that in the order $\mu^N$, gravity on
either brane is indistinguishable from a higher derivative gravity whose
action includes terms of up to the $(N+1)$-th power of curvature
tensors. Noting that the expansion in $\mu$ is in principle an infinite
series, this conjecture indicates that gravity on either brane is
non-local at high energies even at the linearized level. This explains
how the $4$-dimensional description breaks down at high
energies. Physically, the non-locality is due to gravitational and
scalar waves in the bulk.


This paper is organized as follows. In section~\ref{sec:basic-eq} we
summarize the basic equations by reviewing the formulation given in 
ref.~\cite{Mukohyama-Kofman}. In section~\ref{sec:expansion} we perform
the low energy expansion to investigate the system. In
section~\ref{sec:higher-derivative} we review linear perturbations in
$4$-dimensional higher derivative gravity so as to compare it with
gravity in the brane world. In section~\ref{sec:implications} we discuss
some physical implications. Finally, section~\ref{sec:summary} is
devoted to a summary of results.


\section{Basic equations}
\label{sec:basic-eq}

The model and basic equations we shall investigate in this paper are
exactly the same as those in ref.~\cite{Mukohyama-Kofman}. We now
summarize them briefly.

\subsection{Model description and background}

We consider a $5$-dimensional spacetime ${\cal M}$ of the
topology ${\cal M}_4\otimes S^1/Z_2$, where ${\cal M}_4$ represents
$4$-dimensional spacetime. We denote two timelike hypersurfaces
corresponding to fixed points of the $S^1/Z_2$ compactification by
$\Sigma_{\pm}$. Each hypersurface can be considered as the world volume
of a $3$-brane. In order to describe $\Sigma_{\pm}$ we use the
parametric equations
%
\begin{equation}
 \Sigma_{\pm}:\quad x^M = Z^M_{\pm}(y_{\pm}),
\end{equation}
where $x^M$ $(M=0,\cdots,4)$ are $5$-dimensional coordinates in 
${\cal M}$ and each $y_{\pm}$ denotes four parameters $\{y_{\pm}^{\mu}\}$
$(\mu=0,\cdots,3)$. The four parameters play the role of $4$-dimensional
coordinates on each hypersurface. It is notable that the $4$-dimensional
coordinates $y_{\pm}^{\mu}$ are not necessarily a part of the
$5$-dimensional coordinates. Actually, in the following we shall
consider a $4$-dimensional gauge transformation (a $4$-gauge
transformation) on each brane and a $5$-dimensional gauge transformation
(a $5$-gauge transformation) in the bulk independently, where we call
the $5$-dimensional region between $\Sigma_{\pm}$ the bulk
(and we shall denote it by ${\cal M}_b$). In particular, a quantity
invariant under the latter (a $5$-gauge invariant variable) is not
necessarily invariant under the former ($4$-gauge invariant).

We consider a theory described by the action
%
\begin{equation}
 I_{tot} = I_{EH} + I_{\Psi} + I_{matter},
  \label{eqn:total-action}
\end{equation}
where $I_{EH}$ is the $5$-dimensional Einstein-Hilbert action 
%
\begin{equation}
 I_{EH} = \frac{1}{2\kappa_5^2}\int_{\cal M} d^5x\sqrt{-g}R,
\end{equation}
$I_{\Psi}$ is the action of a scalar field $\Psi$
%
\begin{equation}
 I_{\Psi} = -\int_{{\cal M}_b} d^5x\sqrt{-g}
  \left[\frac{1}{2}g^{MN}\partial_M\Psi\partial_N\Psi
   +U(\Psi)\right]
  - \sum_{\sigma=\pm}\int_{\Sigma_{\sigma}}d^4y_{\sigma}
  \sqrt{-q_{\sigma}}V_{\sigma}(\Psi_{\sigma})
  \label{eqn:einstein-action}
\end{equation}
and $I_{matter}$ is the action of matter fields confined on the branes
%
\begin{equation}
 I_{matter} = \sum_{\sigma=\pm}\int_{\Sigma_{\sigma}}d^4y_{\sigma}
     {\cal L}_{\pm}[\bar{q}_{\pm\mu\nu},\mbox{matter}]. 
     \label{eqn:matter-action}
\end{equation}
Here, $\Psi_{\pm}$ and $q_{\pm\mu\nu}$ represent the pullback of $\Psi$
and the induced metric on $\Sigma_{\pm}$, and $\bar{q}_{\pm\mu\nu}$ is
the physical metric on $\Sigma_{\pm}$, which is not necessarily
equivalent to $q_{\pm\mu\nu}$. We assume that the physical metric is
related to the induced metric by a conformal transformation depending on
$\Psi_{\pm}$:
%
\begin{equation}
 \bar{q}_{\pm\mu\nu} = \exp[-\alpha_{\pm}(\Psi_{\pm})]q_{\pm\mu\nu}, 
  \label{eqn:physical-metric}
\end{equation}
where $\alpha_{\pm}$ is a function of $\Psi_{\pm}$, respectively. 
As shown in ref.~\cite{Mukohyama2001b} the variational principle based
on the action (\ref{eqn:total-action}) gives the correct set of
equations of motion, including variations of $Z^M_{\pm}$, $g_{MN}$ and
$\Psi$. It is essential that the region of integration in the Einstein
action (\ref{eqn:einstein-action}) is not ${\cal M}_b$ but ${\cal M}$: 
the integration across $\Sigma_{\pm}$ gives the so called Gibons-Hawking
term correctly.

We consider general perturbations around a background
with $4$-dimensional Poincare symmetry:
%
\begin{eqnarray}
 g_{MN} & = & g^{(0)}_{MN} + \delta g_{MN}, \nonumber\\
 \Psi & = & \Psi^{(0)} + \delta\Psi, \nonumber\\
 Z_{\pm}^M & = & Z_{\pm}^{(0)M} + \delta Z_{\pm}^M, \nonumber\\
 \bar{S}_{\pm\mu\nu} & = & \bar{S}^{(0)}_{\pm\mu\nu} 
 + \delta\bar{S}_{\pm\mu\nu},
\end{eqnarray}
where the background is given by 
%
\begin{eqnarray}
 g^{(0)}_{MN}dx^Mdx^N & = & e^{-2A(w)}(\eta_{\mu\nu}dx^{\mu}dx^{\nu}+dw^2), 
  \nonumber\\
 \Psi^{(0)} & = & \Psi^{(0)}(w), \nonumber\\
 Z_{\pm}^{(0)\mu} & = & y^{\mu}_{\pm}, \nonumber\\
 Z_{\pm}^{(0)w} & = & w_{\pm}, \nonumber\\
 \bar{S}^{(0)}_{\pm\mu\nu} & = & 0,
  \label{eqn:background-ansatz}
\end{eqnarray}
$\{x^{\mu}\}$ ($\mu=0,\cdots,3$) represent first four of $5$-dimensional 
coordinates $\{x^M\}$ ($M=0,\cdots,4$) in ${\cal M}$, $w$ represents the 
fifth coordinate $x^4$, and $w_{\pm}$ ($w_-<w_+$) are constants. Here,
$\bar{S}_{\pm}^{\mu\nu}$ is the physical surface energy momentum tensor
defined by 
%
\begin{equation}
 \bar{S}_{\pm}^{\mu\nu} \equiv \frac{2}{\sqrt{-\bar{q}_{\pm}}}
  \frac{\delta}{\delta\bar{q}_{\pm\mu\nu}}\int_{\Sigma_{\pm}}d^4y_{\pm}
  {\cal L}_{\pm}[\bar{q}_{\pm\mu\nu},\mbox{matter}],
\end{equation}
and we have redefined $V_{\pm}$ and ${\cal L}_{\pm}$ so that
$\bar{S}^{(0)}_{\pm\mu\nu}$ vanishes. Hereafter, we assume that 
the brane at $w=w_+$ is our brane and that there is no excitation of
matter on the other brane ($\bar{S}_-^{\mu\nu}=0$).

The equations of motion for the background are as follows. 
%
\begin{eqnarray}
 3\ddot{A}+ 3\dot{A}^2 & = & \kappa_5^2\dot{\Psi}^{(0)2}, \nonumber\\
 3\ddot{A}-9\dot{A}^2 & = & 2\kappa_5^2e^{-2A}U(\Psi^{(0)})
  \label{eqn:background-bulk-eq}
\end{eqnarray}
and 
%
\begin{eqnarray}
 \left.e^A\dot{A}\right|_{w=w_{\pm}} & = & 
  \mp\frac{1}{6}\kappa_5^2V_{\pm}(\Psi^{(0)}_{\pm})
  \nonumber\\
 \left.e^A\dot{\Psi}^{(0)}\right|_{w=w_{\pm}} & = & 
  \mp\frac{1}{2}V'_{\pm}(\Psi^{(0)}_{\pm}),
  \label{eqn:background-junction}
\end{eqnarray}
where dots denote derivative with respect to $w$ and
$\Psi^{(0)}_{\pm}=\Psi^{(0)}(w_{\pm})$. 
Here, we assumed that $w_-<w_+$ and that the bulk is the region
$w_-<w<w_+$. The induced metric and the physical metric on
$\Sigma_{\pm}$ for the background are 
%
\begin{eqnarray}
 q_{\pm\mu\nu}^{(0)} & = & e^{-2A(w_{\pm})}\eta_{\mu\nu},\nonumber\\
 \bar{q}_{\pm\mu\nu}^{(0)} & = & 
 e^{-\alpha_{\pm}(\Psi^{(0)}_{\pm})}e^{-2A(w_{\pm})}\eta_{\mu\nu}. 
\end{eqnarray}

\subsection{Gauge-invariant variables}
\label{subsec:variables}

Let us now construct gauge-invariant variables from the metric
perturbation $\delta g_{MN}$, the scalar field perturbation
$\delta\Psi$, the brane fluctuation $\delta Z_{\pm}^M$ and the matter
perturbation $\delta\bar{S}_{\pm\mu\nu}$. There are actually two types
of gauge-invariant variables as there are two types of
gauge-transformations: the $5$-dimensional gauge transformation in the
bulk ($5$-gauge transformation) 
%
\begin{equation}
 x^M \to x^M + \bar{\xi}^M(x),
\end{equation}
and the $4$-dimensional gauge transformation on each brane
$\Sigma_{\pm}$ ($4$-gauge transformation)
%
\begin{equation}
 y_{\pm}^{\mu} \to y_{\pm}^{\mu} + \bar{\zeta}_{\pm}^{\mu}(y_{\pm}). 
\end{equation}
As pointed out in ref.~\cite{Mukohyama2000c}, these two kinds of 
gauge-transformations are independent.

For the purpose of construction of gauge-invariant variables we expand
all perturbations by harmonics in $4$-dimensional Minkowski
spacetime. This strategy is convenient since the background has
$4$-dimensional Poincare symmetry and the induced (and physical) metric
on each brane is $4$-dimensional Minkowski metric. In
appendix~\ref{app:harmonics} we define scalar harmonics $Y$, vector
harmonics $V_{(T,L)\mu}$ and tensor harmonics
$T_{(T,LT,LL,Y)\mu\nu}$. By using those harmonics we can expand all
perturbations as
%
\begin{eqnarray}
 \delta g_{MN}dx^Mdx^N & = & 
  ( h_{(T)}T_{(T)\mu\nu} + h_{(LT)}T_{(LT)\mu\nu}
   + h_{(LL)}T_{(LL)\mu\nu} + h_{(Y)}T_{(Y)\mu\nu})dx^{\mu}dx^{\nu}
    \nonumber\\
  & & + 2( h_{(T)w}V_{(T)\mu}+h_{(L)w}V_{(L)\mu})dx^{\mu}dw
   +h_{ww}Ydw^2, \nonumber\\
 \delta\Psi & = & \psi Y, \nonumber\\
 \delta Z_{\pm M}dx^M & = & 
 (z_{\pm(T)}V_{(T)\mu}+z_{\pm(L)}V_{(L)\mu})dx^{\mu}
 + z_{\pm w}Ydw,
 \label{eqn:harmonic-expansion}
\end{eqnarray}
and
%
\begin{equation}
 \delta\bar{S}_{\pm\mu\nu} =  
  \bar{\tau}_{\pm(T)}T_{(T)\mu\nu} + \bar{\tau}_{\pm(LT)}T_{(LT)\mu\nu} 
   + \bar{\tau}_{\pm(LL)}T_{(LL)\mu\nu} + \bar{\tau}_{\pm(Y)}T_{(Y)\mu\nu}, 
\end{equation}
where we omitted dependence of harmonics and the corresponding
coefficients on the $4$-dimensional momentum $k_{\mu}$ and the
integration with respect to $k_{\mu}$. The $k$-dependent Fourier
coefficients $h_{(T,LT,LL,Y)}$, $h_{(T,L)w}$, $h_{ww}$, and $\psi$ are
functions of the fifth coordinate $w$ only, and the other coefficients
$z_{\pm(T,L)}$, $z_{\pm w}$, and $\tau_{\pm(T,LT,LL,Y)}$ are constants.

Now we can analyze $5$-gauge transformation of the coefficients of the
harmonic expansion and construct $5$-gauge-invariant variables, or those
linear combinations of perturbations that are invariant under the
$5$-gauge transformation. The result is 
%
\begin{eqnarray}
 F_{(T)} & = & h_{(T)}, \nonumber\\
 F_w & = & h_{(T)w} - e^{-2A}(e^{2A}h_{(LT)})^{\cdot}, \nonumber\\ 
 F & = & h_{(Y)} + 2 \dot{A}X_w 
  + \frac{1}{2}\eta^{\mu\nu}k_{\mu}k_{\nu}h_{(LL)}, \nonumber\\
 F_{ww} & = & h_{ww} - 2e^{-A}(e^AX_w)^{\cdot}, \nonumber\\
 \varphi & = & \psi - e^{2A}\dot{\Psi}^{(0)}X_w,\nonumber\\
 \phi_{\pm(T)} & = & z_{\pm(T)} + \left.h_{(LT)}\right|_{w=w_{\pm}}, 
  \nonumber\\
 \phi_{\pm(L)} & = & z_{\pm(L)} + \left.h_{(LL)}\right|_{w=w_{\pm}}, 
  \nonumber\\
 \phi_{\pm w} & = & z_{\pm w} + \left.X_w\right|_{w=w_{\pm}}, 
\end{eqnarray}
where $X_w=h_{(L)w}-e^{-2A}(e^{2A}h_{(LL)})^{\cdot}$. They form a
maximal set of independent $5$-gauge invariant variables constructed
from the metric perturbation $\delta g_{MN}$, the scalar field
perturbation $\delta\Psi$ and the brane fluctuation $\delta Z_{\pm}^M$. 
(The matter perturbation $\delta\bar{S}_{\pm\mu\nu}$ is not included
here since it is a $4$-dimensional object.)

We can also analyze $4$-gauge transformation and construct
$4$-gauge-invariant variables, or those linear combinations of
perturbations that are invariant under the $4$-gauge transformation. For
this purpose we first need to obtain expressions of various
$4$-dimensional quantities in terms of the $5$-dimensional quantities
$g_{MN}$, $\Psi$ and $Z_{\pm}^M$. What we need are the physical metric
$\bar{q}_{\pm\mu\nu}$ defined by (\ref{eqn:physical-metric}), the
extrinsic curvature $K_{\pm\mu\nu}$, the pull back $\Psi_{\pm}$ of the
scalar field $\Psi$, and the normal derivative
$\partial_{\perp}\Psi_{\pm}$ of $\Psi$ on $\Sigma_{\pm}$:
%
\begin{eqnarray}
 \bar{q}_{\pm\mu\nu}(y_{\pm}) & = & \exp[-\alpha_{\pm}(\Psi_{\pm})]
  \left.e^M_{\pm\mu}e^N_{\pm\nu}g_{MN}\right|_{x=Z_{\pm}(y_{\pm})},
  \nonumber\\
 K_{\pm\mu\nu}(y_{\pm}) & = &
  \frac{1}{2}\left.e^M_{\pm\mu}e^N_{\pm\nu}
        {\cal L}_{n_{\pm}}g_{MN}\right|_{x=Z_{\pm}(y_{\pm})},
  \nonumber\\
 \Psi_{\pm}(y_{\pm}) & = & \left.\Psi\right|_{x=Z_{\pm}(y_{\pm})},
  \nonumber\\
 \partial_{\perp}\Psi_{\pm} & = & 
  \left.n_{\pm}^M\partial_M\Psi\right|_{x=Z_{\pm}(y_{\pm})},
\end{eqnarray}
where 
%
\begin{equation}
 e_{\pm\mu}^M = \frac{\partial Z_{\pm}^M}{\partial y_{\pm}^{\mu}},
\end{equation}
and ${\cal L}$ represents a $5$-dimensional Lie derivative. Using the
harmonic expansions (\ref{eqn:harmonic-expansion}), we can obtain the
corresponding harmonic expansion of $\delta\bar{q}_{\pm\mu\nu}$,
$\delta K_{\pm\mu\nu}$, $\delta\Psi_{\pm}$ and
$\delta\partial_{\perp}\Psi_{\pm}$. The result is 
%
\begin{eqnarray}
 \delta\bar{q}_{\pm\mu\nu} & = & 
   \bar{\sigma}_{\pm(T)}T_{(T)\mu\nu} 
   + \bar{\sigma}_{\pm(LT)}T_{(LT)\mu\nu}
   + \bar{\sigma}_{\pm(LL)}T_{(LL)\mu\nu} 
   + \bar{\sigma}_{\pm(Y)}T_{(Y)\mu\nu}, 
    \nonumber\\
 \delta \tilde{K}_{\pm\mu\nu} & = & 
   k_{\pm(T)}T_{(T)\mu\nu} + k_{\pm(LT)}T_{(LT)\mu\nu}
   + k_{\pm(LL)}T_{(LL)\mu\nu} + k_{\pm(Y)}T_{(Y)\mu\nu},
   \nonumber\\
 \delta\Psi_{\pm} & = & \psi_{\pm}Y, \nonumber\\
 \delta\partial_{\perp}\Psi_{\pm} & = & \psi_{\pm\perp}Y, 
\end{eqnarray}
where $\delta\tilde{K}_{\pm\mu\nu}$ is defined by
%
\begin{equation}
 \delta\tilde{K}_{\pm\mu\nu} \equiv 
        \delta K_{\pm\mu\nu} - \frac{1}{2}
        (K^{(0)\rho}_{\pm\mu}\delta q_{\pm\rho\nu}
        +K^{(0)\rho}_{\pm\nu}\delta q_{\pm\rho\mu}), 
\end{equation}
and $k$-dependent Fourier coefficients are
%
\begin{eqnarray}
 \bar{\sigma}_{\pm(T)} & = & e^{-\alpha^{(0)}_{\pm}}F_{(T)},
  \nonumber\\
 \bar{\sigma}_{\pm(LT)} & = & e^{-\alpha^{(0)}_{\pm}}\phi_{\pm(T)},
  \nonumber\\
 \bar{\sigma}_{\pm(LL)} & = & e^{-\alpha^{(0)}_{\pm}}\phi_{\pm(L)},
  \nonumber\\
 \bar{\sigma}_{\pm(Y)} & = & e^{-\alpha^{(0)}_{\pm}}
  \left[ F - 2\dot{A}\phi_{\pm w} 
   - \frac{1}{2}\eta^{\mu\nu}k_{\mu}k_{\nu}\phi_{\pm(L)}
  -{\alpha'}_{\pm}^{(0)}
  (e^{-2A}\varphi+\phi_{\pm w}\dot{\Psi}^{(0)})\right],
  \nonumber\\
 k_{\pm(T)} & = & \mp\frac{1}{2}e^{-A}(e^{2A}F_{(T)})^{\cdot},
  \nonumber\\
 k_{\pm(LT)} & = & \pm\frac{1}{2}e^AF_w,\nonumber\\
 k_{\pm(LL)} & = & \pm\frac{1}{2}e^A\phi_{\pm w},\nonumber\\
 k_{\pm(Y)} & = & \mp\frac{1}{2}
  \left\{e^{-A}(e^{2A}F)^{\cdot}+(e^A)^{\cdot}F_{ww}
   +\left[\frac{1}{2}e^A\eta^{\mu\nu}k_{\mu}k_{\nu}-2(e^A)^{\cdot\cdot}
   \right]\phi_{\pm }\right\}, \nonumber\\
 \psi_{\pm} & = & \varphi + e^{2A}\dot{\Psi}^{(0)}\phi_{\pm w},\nonumber\\
 \psi_{\pm\perp} & = & \mp\frac{1}{2}
  \left[-e^{3A}\dot{\Psi}^{(0)}F_{ww}+2e^A\dot{\varphi}
   +2e^{2A}(e^A\dot{\Psi}^{(0)})^{\cdot}\phi_{\pm w}\right].
   \label{eqn:sigma-k}
\end{eqnarray}
Here, the right hand sides of (\ref{eqn:sigma-k}) are evaluated at
$w=w_{\pm}$, respectively, and have been written in terms of $5$-gauge
invariant variables~\footnote{The reason why they can be expressed in
terms of $5$-gauge-invariant variables only is that they by themselves
are $5$-gauge-invariant~\cite{Mukohyama2000b}. This fact illustrates
that $4$-gauge transformation is not a part of $5$-gauge transformation
and that these two kinds of gauge transformations are independent.}. The
matter perturbation $\delta\bar{S}_{\pm\mu\nu}$ on each brane can also
be expanded by harmonics as
%
\begin{equation}
 \delta\bar{S}_{\pm\mu\nu} =  
  \bar{\tau}_{\pm(T)}T_{(T)\mu\nu} + \bar{\tau}_{\pm(LT)}T_{(LT)\mu\nu} 
   + \bar{\tau}_{\pm(LL)}T_{(LL)\mu\nu} + \bar{\tau}_{\pm(Y)}T_{(Y)\mu\nu}.
\end{equation}

We can now analyze $4$-gauge transformation of coefficients the
harmonic expansion and construct the following $4$-gauge invariant
variables from the physical metric perturbation
$\delta\bar{q}_{\pm\mu\nu}$. 
%
\begin{eqnarray}
 \bar{f}_{\pm(T)} & = & \bar{\sigma}_{\pm(T)} 
  = e^{-\alpha^{(0)}_{\pm}}F_{(T)},\nonumber\\
 \bar{f}_{\pm} & = & \bar{\sigma}_{\pm(Y)} 
  + \frac{1}{2}\eta^{\mu\nu}k_{\mu}k_{\nu}\bar{\sigma}_{\pm(LL)}
  = e^{-\alpha^{(0)}_{\pm}}
  \left[ F - 2\dot{A}\phi_{\pm w} 
  -{\alpha'_{\pm}}^{(0)}
  (e^{-2A}\varphi+\phi_{\pm w}\dot{\Psi}^{(0)})\right]. 
  \label{eqn:fT-f}
\end{eqnarray}
It is easily shown that $k_{\pm(T,LT,LL,Y)}$, $\psi_{\pm}$, 
$\psi_{\pm\perp}$ and $\bar{\tau}_{(T,LT,LL,Y)}$ are invariant under the
$4$-gauge transformation. Moreover, they are at the same time $5$-gauge
invariant since all except for $\bar{\tau}_{(T,LT,LL,Y)}$ are written
in terms of $5$-gauge invariant variables and
$\delta\bar{S}_{\pm\mu\nu}$ is a $4$-dimensional object. Hence, we have
the set ($\bar{f}_{\pm(T)}$, $\bar{f}_{\pm}$, $k_{\pm(T,LT,LL,Y)}$,
$\psi_{\pm}$, $\psi_{\pm\perp}$, $\bar{\tau}_{(T,LT,LL,Y)}$) of
doubly-gauge invariant variables.

From the point of view of observers on each brane, all observable
quantities must be doubly-gauge invariant. However, all doubly-gauge
invariant variables are not necessarily observable quantities. They can 
observe physical metric perturbation ($\bar{f}_{\pm(T)}$,
$\bar{f}_{\pm}$) and matter perturbation $\bar{\tau}_{(T,LT,LL,Y)}$
only. The remaining doubly-gauge invariant variables
($k_{\pm(T,LT,LL,Y)}$, $\psi_{\pm}$, $\psi_{\pm\perp}$) shall be used to
write down junction conditions of $5$-dimensional quantities in a
doubly-gauge invariant way.

Our remaining task in this section is to give $5$-gauge invariant
equations in the bulk and doubly-gauge invariant junction conditions on
each brane. Our final aim in this paper is to seek doubly-gauge
invariant equations governing the physical metric perturbations and
matter perturbations on our brane and to compare the resulting
equations with the corresponding equations in $4$-dimensional higher
derivative gravity.

Since there are many coefficients in the above harmonic expansions, let 
us divide these into three classes. The first class is the scalar
perturbations and consists of coefficients of $Y$, $V_{(L)\mu}$, 
$T_{(LL)\mu\nu}$ and $T_{(Y)\mu\nu}$. The second is the vector
perturbations and consists of coefficients of $V_{(T)\mu}$ and
$T_{(LT)\mu\nu}$. The last is the tensor perturbations and consists of
coefficients of $T_{(T)\mu\nu}$. Perturbations in different classes are
decoupled from each other at the linearized level. Hence, in the
following we analyze perturbations in each class
separately. Decomposition into scalar, vector and tensor modes 
is commonly used in cosmology. However, usually in cosmology we use
scalar, vector and tensor representations of the isometry group
related to the symmetry of $3$-dimensional space. Meanwhile, here we
will use scalar, vector and tensor representations of the isometry group
of $4$-dimensional space-time.

In ref.~\cite{Mukohyama-Kofman} it was shown that vector type
perturbations vanish unless matter fields on the hidden brane are 
excited. In this paper we assume that there is no matter excitation on
the hidden brane and, thus, we shall consider scalar and tensor type
perturbations only.

\subsection{Scalar perturbations}

For scalar perturbations, we have three $5$-gauge invariant variables
from metric perturbation and scalar field perturbation in the bulk: $F$,
$F_{ww}$ and $\varphi$. The first two are from metric perturbation and
the last one is from scalar field perturbation. The Einstein equation
leads to two relations among them 
%
\begin{eqnarray}
 F_{ww} & = & -2F, \nonumber\\
 \varphi & = & -\frac{3e^{2A}}{2\kappa_5^2\dot{\Psi}^{(0)}}\dot{F}, 
\end{eqnarray}
and a wave equation
%
\begin{equation}
 \frac{\ddot{A}+\dot{A}^2}{\dot{A}^2e^{3A}}
  \left[\frac{\dot{A}^2e^{3A}}{\ddot{A}+\dot{A}^2}
   \left(\frac{F}{\dot{A}e^A}\right)^{\cdot}\right]^{\cdot}
   -\eta^{\mu\nu}k_{\mu}k_{\nu}\left(\frac{F}{\dot{A}e^A}\right) 
   = 0, \label{eqn:scalar-bulkeq}
\end{equation}
where a dot denotes derivative with respect to $w$ and $k_{\mu}$ is the
$4$-dimensional momentum in the coordinate $y^{\mu}$. Throughout this
paper we consider modes with $k_{\mu}\ne 0$ for scalar perturbations
since a scalar mode with $k_{\mu}=0$ preserves $4$-dimensional Poincare 
symmetry and, thus, represents just a change of the background within
the ansatz (\ref{eqn:background-ansatz}). Of course we shall consider
modes with $\eta^{\mu\nu}k_{\mu}k_{\nu}=0$ as long as 
$k_{\mu}\ne 0$.

The boundary condition at $w=w_{\pm}$, respectively, is
given by the junction condition for the scalar field as 
%
\begin{equation}
 \mp\left[-e^{3A}\dot{\Psi}^{(0)}F_{ww}+2e^A\dot{\varphi}
   +2e^{2A}(e^A\dot{\Psi}^{(0)})^{\cdot}\phi_{\pm w}\right]
 = {V''_{\pm}}^{(0)}(\varphi+e^{2A}\dot{\Psi}^{(0)}\phi_{\pm w})
 + 2{\alpha'_{\pm}}^{(0)}e^{2A}\bar{\tau}_{\pm(Y)},
 \label{eqn:scalar-matching-pre}
\end{equation}
where $\bar{\tau}_{\pm(Y)}$ and $\phi_{\pm w}$ are doubly gauge
invariant variables constructed from matter on $\Sigma_{\pm}$ and 
perturbation of the position of $\Sigma_{\pm}$, respectively, and
satisfy the following equations derived from the perturbed Israel's
junction condition~\footnote{For a mode with $k_{\mu}=0$ we do not have
the second equation of (\ref{eqn:scalar-junction}) since there is no
tensor harmonics of the type $(LL)$ for $k_{\mu}=0$. See appendix
\ref{app:harmonics} for definition and properties of harmonics.}. 
%
\begin{eqnarray}
 2\bar{\tau}_{\pm(Y)} & = & 
  3\eta^{\mu\nu}k_{\mu}k_{\nu}\bar{\tau}_{\pm(LL)}, \nonumber\\
 \phi_{\pm w} & = & 
  \mp\kappa_5^2e^{-A-\alpha^{(0)}_{\pm}}\bar{\tau}_{\pm(LL)}. 
  \label{eqn:scalar-junction}
\end{eqnarray}
Here,
$\alpha^{(0)}_{\pm}=\left.\alpha(\Psi^{(0)})\right|_{w=w_{\pm}}$, 
${\alpha'}_{\pm}^{(0)}=\left.\alpha'(\Psi^{(0)})\right|_{w=w_{\pm}}$ 
and $\bar{\tau}_{\pm(LL)}$ is another doubly gauge invariant variable 
constructed from matter on $\Sigma_{\pm}$. The first of
(\ref{eqn:scalar-junction}) is nothing but the conservation equation of
matter stress energy tensor $\bar{S}_{\pm\mu\nu}$ on each brane. The
second equation is the $(LL)$-component of the Israel's junction
condition and relates the perturbation of the brane position and the
matter perturbation. The boundary condition
(\ref{eqn:scalar-matching-pre}) at $w=w_{\pm}$, respectively, can be 
rewritten to the following form by eliminating $\varphi$,
$\bar{\tau}_{\pm(Y)}$ and $\phi_{\pm w}$, and using the wave equation
(\ref{eqn:scalar-bulkeq}). 
%
\begin{equation}
 C_{\pm}\left[ \dot{F} \pm 2\kappa_5^2e^{-A-\alpha_{\pm}^{(0)}}
   (\ddot{A}+\dot{A}^2)\bar{\tau}_{\pm(LL)}\right]
 +\eta^{\mu\nu}k_{\mu}k_{\nu}\dot{\Psi}^{(0)}
 \left[ F\mp \kappa_5^2{\alpha'}_{\pm}^{(0)}
  e^{-A-\alpha_{\pm}^{(0)}}\dot{\Psi}^{(0)}
  \bar{\tau}_{\pm(LL)}\right] =0, \label{eqn:scalar-matching}
\end{equation}
where
%
\begin{equation}
 C_{\pm} = \left.\ddot{\Psi}^{(0)} + \dot{A}\dot{\Psi}^{(0)}
  \pm\frac{1}{2}e^{-A}\dot{\Psi}^{(0)}V''(\Psi^{(0)})\right|_{w=w_{\pm}}.
\end{equation}

Finally, the doubly gauge invariant perturbation of the physical metric
$\bar{q}_{\pm\mu\nu}$ is expressed as
%
\begin{equation}
 \bar{f}_{\pm} = \left.e^{-\alpha^{(0)}_{\pm}}
  \left[ F - 2\dot{A}\phi_{\pm w} 
  -{\alpha'_{\pm}}^{(0)}
  (e^{-2A}\varphi+\phi_{\pm w}\dot{\Psi}^{(0)})\right]
  \right|_{w=w_{\pm}}. 
\end{equation}
In the next section we shall assume that 
$\dot{\Psi}^{(0)}(w_{\pm})C_{\pm}\ne 0$ to show 
the recovery of higher derivative gravity on a brane. In this case, the
expression of $\bar{f}_{\pm}$ can be rewritten to the following form by
eliminating $\varphi$ and $\phi_{\pm w}$, and using the boundary
condition (\ref{eqn:scalar-matching}). 
%
\begin{equation}
 e^{\alpha^{(0)}_{\pm}}\bar{f}_{\pm} = \left.
 F\pm2\kappa_5^2\dot{A}e^{-A-\alpha_{\pm}^{(0)}}\bar{\tau}_{\pm(LL)}
   -\eta^{\mu\nu}k_{\mu}k_{\nu}
   \frac{3{\alpha'}_{\pm}^{(0)}}{2\kappa_5^2C_{\pm}}
   \left[ F\mp \kappa_5^2{\alpha'}_{\pm}^{(0)}
  e^{-A-\alpha_{\pm}^{(0)}}\dot{\Psi}^{(0)}
  \bar{\tau}_{\pm(LL)}\right]\right|_{w=w_{\pm}}. 
  \label{eqn:bar-f}
\end{equation}

Hereafter, we consider $\Sigma_+$ as our brane and $\Sigma_-$ as the
hidden brane, and assume that there is no matter excitations on the 
hidden brane. Hence, we put $\bar{\tau}_{-(LL,Y)}=0$.

\subsection{Tensor perturbations}

For tensor perturbations we have only one $5$-gauge invariant variable
constructed from metric perturbation in the bulk: $F_{(T)}$. The
Einstein equation leads to the wave equation 
%
\begin{equation}
 e^A\left[ e^{-3A}(e^{2A}F_{(T)})^{\cdot}\right]^{\cdot}
  -\eta^{\mu\nu}k_{\mu}k_{\nu}F_{(T)} = 0,
  \label{eqn:tensor-bulkeq}
\end{equation}
where $k^{\mu}$ is the $4$-dimensional momentum in the coordinate
$y^{\mu}$.

The Israel junction condition leads to the following boundary
condition at $w=w_{\pm}$, respectively.
%
\begin{equation}
 (e^{2A}F_{(T)})^{\cdot} = \pm\kappa_5^2e^{A_{\pm}-\alpha^{(0)}_{\pm}}
  \bar{\tau}_{\pm(T)}, \label{eqn:bound-cond-tensor}
\end{equation}
where $\bar{\tau}_{\pm(T)}$ is a doubly gauge invariant variable 
constructed from matter on $\Sigma_{\pm}$.

Finally, the doubly gauge invariant perturbation of the physical metric
$\bar{q}_{\pm\mu\nu}$ is expressed as
%
\begin{equation}
 \bar{f}_{\pm(T)} = 
  \left.e^{-\alpha^{(0)}_{\pm}}F_{(T)}\right|_{w=w_{\pm}}. 
\end{equation}

Hereafter, since we assumed that there is no matter excitations on the
hidden brane, we put $\bar{\tau}_{-(T)}=0$.


\section{Low energy expansion}
\label{sec:expansion}

As already stated in the third-to-the-last paragraph of subsection
\ref{subsec:variables}, our aim in this paper is to seek doubly-gauge
invariant equations governing the physical metric perturbations and
matter perturbations on our brane and to compare the resulting
equations with the corresponding equations in $4$-dimensional higher
derivative gravity. Since we have only two gauge-invariant metric
perturbations $\bar{f}_+$ and $\bar{f}_{+(T)}$, and essentially two 
gauge-invariant matter perturbations $\bar{\tau}_{+(LL)}$ and
$\bar{\tau}_{+(T)}$ on our brane ($\bar{\tau}_{+(Y)}$ is related to
$\bar{\tau}_{+(LL)}$ by the conservation equation), we expect the
following form of the equations on the brane. 
%
\begin{eqnarray}
 C_{(S)}\bar{f}_+ & = & \bar{\tau}_{+(LL)}, \nonumber\\ 
 \bar{q}_+^{(0)\mu\nu}k_{\mu}k_{\nu}C_{(T)}\bar{f}_{+(T)} 
  & = & \bar{\tau}_{+(T)},
  \label{eqn:expectation}
\end{eqnarray}
where $C_{(S)}$ and $C_{(T)}$ are functions of
$\bar{q}_+^{(0)\mu\nu}k_{\mu}k_{\nu}$ and 
$\bar{q}_+^{(0)\mu\nu}=\Omega_+^{-2}\eta^{\mu\nu}$ is the inverse of the
unperturbed physical metric $\bar{q}_{+\mu\nu}^{(0)}$. The reason why we
expect the linear dependence of metric perturbations on matter
perturbations is that boundary conditions summarized in the previous
section are linear in the matter perturbations. The reason why 
$\bar{q}_+^{(0)\mu\nu}k_{\mu}k_{\nu}$ was put in front of
$\bar{f}_{+(T)}$ is that we expect $4$-dimensional gravitons on our
brane (non-vanishing $\bar{f}_{+(T)}$ with
$\bar{q}_+^{(0)\mu\nu}k_{\mu}k_{\nu}=0$ and $\bar{\tau}_{+(T)}=0$). What
is important here is that the functions $C_{(S,T)}$ completely
characterize the effective theory of weak gravity on our
brane~\footnote{If one likes, one can restore gauge fixed
equations for any gauge choices from the functions $C_{(S,T)}$ only. For
example, see (\ref{eqn:result-gauge-fixed}).}. Since we are dealing with
gauge-invariant variables only, there is no ambiguity of gauge freedom
when we compare the result with the corresponding equations in
$4$-dimensional higher derivative gravity. Namely, we only have to
compare functions $C_{(S,T)}$ of $\bar{q}_+^{(0)\mu\nu}k_{\mu}k_{\nu}$
with the corresponding functions of momentum squared in the Fourier
transformed, linearized $4$-dimensional higher derivative gravity.

In this section we expand the basic equations summarized in the previous
section by the parameter $\mu=l^2\eta^{\mu\nu}k_{\mu}k_{\nu}$ and solve
them iteratively, where $l$ is the characteristic length scale of the
model which we shall determine by comparing the results of order $O(1)$
and $O(\mu)$. The purpose of the $\mu$-expansion is to analyze the
behavior of the functions $C_{(S)}$ and $C_{(T)}$ near $\mu=0$. Namely,
we shall seek first few coefficients $C_{(S,T)}^{[i]}$ ($i=0,1,\cdots$)
of the expansion 
%
\begin{equation}
 C_{(S,T)} = \sum_{i=0}^{\infty}\mu^iC_{(S,T)}^{[i]}. 
  \label{eqn:expansion-C}
\end{equation}
Since the $4$-dimensional physical energy scale $m_+$ on $\Sigma_+$ is
given by $m_+^2=-\bar{q}_+^{(0)\mu\nu}k_{\mu}k_{\nu} 
=-\mu l^{-2}e^{\alpha_+^{(0)}+2A_+}$, the expansion in
$\mu$ is nothing but the low energy expansion. Hence, the first few
coefficients $C_{(S,T)}^{[i]}$ ($i=0,1,\cdots$) of the expansion 
(\ref{eqn:expansion-C}) determine the low energy behavior of weak
gravity on our brane. We expect that $C_{(S,T)}^{[0]}$ give
$4$-dimensional Einstein gravity, that $C_{(S,T)}^{[1]}$ give
curvature-squared corrections to the Einstein gravity, and so on. The
length scale $l$ gives the energy scale $l^{-1}e^{\alpha_+^{(0)}/2+A_+}$
below which we can trust the $\mu$-expansion. Nonetheless, we can defer
the determination of $l$ until we will obtain the results of order
$O(\mu)$ since $l$ can be eliminated from all formal results in each
order of the $\mu$-expansion. Meanwhile, we shall keep it in
intermediate calculations in order to make the expansion parameter
dimensionless.

First, by expanding $F$ and $F_{(T)}$ as
%
\begin{eqnarray}
 F(w) & = & \sum_{i=0}^{\infty}\mu^iF^{[i]}(w), 
  \nonumber\\
 F_{(T)}(w) & = & \sum_{i=0}^{\infty}\mu^iF_{(T)}^{[i]}(w), 
\end{eqnarray}
we can solve the wave equations (\ref{eqn:scalar-bulkeq}) and
(\ref{eqn:tensor-bulkeq}) order by order. The result is 
%
\begin{eqnarray}
 F^{[i]}(w) & = & \dot{A}(w)e^{A(w)}\left[C^{[i]}_1
 + C^{[i]}_2\int_{w_-}^wdw'\frac{\ddot{A}(w')+\dot{A}(w')^2}
 {\dot{A}(w')^2}e^{-3A(w')}\right. \nonumber\\
 & & \left.+l^{-2}\int_{w_-}^wdw'\frac{\ddot{A}(w')+\dot{A}(w')^2}
 {\dot{A}(w')^2}e^{-3A(w')}\int_{w_-}^{w'}dw''
 \frac{\dot{A}(w'')e^{2A(w'')}}{\ddot{A}(w'')+\dot{A}(w'')^2}
 F^{[i-1]}(w'')\right], \nonumber\\
 F_{(T)}^{[i]}(w) & = & 
  e^{-2A(w)}\left[D^{[i]}_1\int_{w_-}^wdw'e^{3A(w')} + D^{[i]}_2
	\right.     \nonumber\\
  & & \left.+ l^{-2}\int_{w_-}^wdw'e^{3A(w')}
  \int_{w_-}^{w'}dw''e^{-A(w'')}F_{(T)}^{[i-1]}(w'')\right],
 \end{eqnarray}
with $F^{[-1]}=F^{[-1]}_{(T)}=0$, where $C^{[i]}_{1,2}$ and
$D^{[i]}_{1,2}$ are constants.

Next, let us analyze the boundary condition. For this purpose, we expand
$\bar{\tau}_{+(LL,T)}$ as 
%
\begin{eqnarray}
 \bar{\tau}_{+(LL)} & = & 
  \sum_{i=0}^{\infty}\mu^i\bar{\tau}_{+(LL)}^{[i]}, \nonumber\\
 \bar{\tau}_{+(T)} & = & 
  \sum_{i=0}^{\infty}\mu^i\bar{\tau}_{+(T)}^{[i]},
\end{eqnarray}
and put $\bar{\tau}_{-(LL)}=\bar{\tau}_{-(T)}=0$ since we assumed that
there is no matter excitation on the hidden brane. The boundary
condition (\ref{eqn:scalar-matching}) for scalar perturbations can be
rewritten as 
%
\begin{eqnarray}
 C^{[i]}_1 + B_+ C^{[i]}_2  & = & 
  -l^{-2}X^{[i-1]}_+ 
  -2\kappa_5^2 e^{-2A_+-\alpha_+^{(0)}}\bar{\tau}_{+(LL)}^{[i]},
  \nonumber\\
 C^{[i]}_1 + B_- C^{[i]}_2  & = & 
  -l^{-2}X^{[i-1]}_-,
\end{eqnarray}
where 
%
\begin{eqnarray}
 B_+ & = & \int_{w_-}^{w_+}dw
  \frac{\dot{A}^2(w)+\ddot{A}(w)}{\dot{A}^2(w)}e^{-3A(w)}
  +\frac{e^{-3A_+}}{\dot{A}_+}, \nonumber\\
 B_- & = & \frac{e^{-3A_-}}{\dot{A}_-}, \nonumber\\
 X^{[i]}_+ & = & \int_{w_-}^{w_+}dw\frac{\ddot{A}(w)+\dot{A}(w)^2}
 {\dot{A}(w)^2}e^{-3A(w)}\int_{w_-}^{w}dw'
 \frac{\dot{A}(w')e^{2A(w')}}{\ddot{A}(w')+\dot{A}(w')^2}
 F^{[i]}(w')   \nonumber\\
 & & +\frac{e^{-3A_+}}{\dot{A}_+}\int_{w_-}^{w_+}dw'
 \frac{\dot{A}(w')e^{2A(w')}}{\ddot{A}(w')+\dot{A}(w')^2}
 F^{[i]}(w')   \nonumber\\
 & & +\frac{3e^{-A_+}}{\kappa_5^2\dot{\Psi}^{(0)}_+C_+}
  \left[F^{[i]}(w_+)-\kappa_5^2{\alpha'}_+^{(0)}
  e^{-A_+-\alpha_+^{(0)}}\dot{\Psi}^{(0)}_+
  \bar{\tau}_{+(LL)}^{[i]}\right], \nonumber\\
 X^{[i]}_- & = & \frac{3e^{-A_-}}{\kappa_5^2\dot{\Psi}^{(0)}_-C_-}
  F^{[i]}(w_-),
\end{eqnarray}
and $X_{\pm}^{[-1]}=0$. We have assumed that
$\dot{\Psi}_{\pm}^{(0)}C_{\pm}\ne 0$ and that $\dot{A}_{\pm}\ne 0$. 
Here, in order to rewrite the last term of $X_+^{[i]}$ and $X_-^{[i]}$,
we have used the background equation (\ref{eqn:background-bulk-eq}). 
We have adopted the following abbreviation: $A_{\pm}=A(w_{\pm})$,
$\dot{A}_{\pm}=\dot{A}(w_{\pm})$, $\ddot{A}_{\pm}=\ddot{A}(w_{\pm})$ and 
$\dot{\Psi}^{(0)}_{\pm}=\dot{\Psi}^{(0)}(w_{\pm})$. 
The boundary condition (\ref{eqn:bound-cond-tensor}) for tensor
perturbations can be easily solved to give 
%
\begin{eqnarray}
 \bar{\tau}_{+(T)}^{[0]} & = & 0, \nonumber\\
 D^{[i]}_1 & = & 0, \nonumber\\
 D^{[i]}_2 & = & 
  \frac{1}{\int_{w_-}^{w_+}dwe^{-3A(w)}}
  \left[\kappa_5^2l^2e^{-2A_+-\alpha_+^{(0)}}\bar{\tau}_{+(T)}^{[i+1]} 
  - l^{-2}Y^{[i-1]}\right],
\end{eqnarray}
where
%
\begin{equation}
 Y^{[i]} = \int_{w_-}^{w_+}dwe^{-3A(w)}\int_{w_-}^wdw'e^{3A(w')}
  \int_{w_-}^{w'}dw''e^{-A(w'')}F_{(T)}^{[i]}(w''),
\end{equation}
and $Y^{[-1]}=0$.

Thirdly, the doubly gauge invariant variables $\bar{f}_+$ and
$\bar{f}_{+(T)}$ corresponding to perturbation of the physical metric
$\bar{q}_{+\mu\nu}$ on $\Sigma_+$ are expanded as
%
\begin{eqnarray}
 \bar{f}_+ & = & \sum_{i=0}^{\infty}\mu^i\bar{f}_+^{[i]}, 
  \nonumber\\
 \bar{f}_{+(T)} & = & \sum_{i=0}^{\infty}\mu^i\bar{f}_{+(T)}^{[i]}, 
\end{eqnarray}
where the expansion coefficients are given by
%
\begin{eqnarray}
 e^{\alpha^{(0)}_+}\bar{f}_+^{[i]} & = &
 F^{[i]}(w_+)+2\kappa_5^2\dot{A}_+e^{-A_+-\alpha_+^{(0)}}
 \bar{\tau}_{+(LL)}^{[i]}
   -\frac{3{\alpha'}_+^{(0)}}{2\kappa_5^2l^2C_+}
   \left[ F^{[i-1]}(w_+) - \kappa_5^2{\alpha'}_+^{(0)}
  e^{-A_+-\alpha_+^{(0)}}\dot{\Psi}^{(0)}_+
  \bar{\tau}_{+(LL)}^{[i-1]}\right],
  \nonumber\\
 e^{\alpha^{(0)}_+}\bar{f}_{+(T)}^{[i]} & = & F_{(T)}^{[i]}(w_+),
\end{eqnarray}
with $F^{[-1]}=\bar{\tau}^{[-1]}_{+[LL]}=0$.

Now let us summarize first few terms in the $\mu$-expansion. 
%
\begin{eqnarray}
 \bar{f}_+^{[0]} & = & 
  -16\pi G_{N+}\Omega_+^2\bar{\tau}_{+(LL)}^{[0]}, 
  \nonumber\\
 \bar{f}_+^{[1]} & = & 
  -16\pi G_{N+}\left[\Omega_+^2\bar{\tau}_{+(LL)}^{[1]} 
		- l^{-2}l_{S+}^2\bar{\tau}_{+(LL)}^{[0]}\right], 
  \nonumber\\
 \bar{\tau}_{+(T)}^{[0]} & = & 0, \nonumber\\
 \bar{f}_{+(T)}^{[0]} & = & 
  16\pi G_{N+}\Omega_+^2l^2\bar{\tau}_{+(T)}^{[1]}, \nonumber\\
 \bar{f}_{+(T)}^{[1]}
  & = & 16\pi G_{N+}
  \left[\Omega_+^2l^2\bar{\tau}_{+(T)}^{[2]} 
   +l_{T+}^2\bar{\tau}_{+(T)}^{[1]}\right],
\end{eqnarray}
where 
%
\begin{eqnarray}
 \Omega_+^2 & = & e^{-2A_+-\alpha^{(0)}_+}, \nonumber\\
 16\pi G_{N+}  & = & 
  \frac{\kappa_5^2\Omega_+^2}
  {\int^{w_+}_{w_-}dwe^{-3A(w)}},\nonumber\\
 l_{S+}^2\times\Omega_+^{-2}\int_{w_-}^{w_+}dwe^{-3A(w)} & = & 
  \frac{1}{2}\int_{w_-}^{w_+}dw\frac{e^{-3A(w)}}{\ddot{A}(w)+\dot{A}(w)^2}
  + 2\int_{w_-}^{w_+}dw\frac{\dot{A}(w)}{\ddot{A}(w)+\dot{A}(w)^2}
  \int_{w_-}^wdw'e^{-3A(w')} 
  \nonumber\\
 & & +2\int_{w_-}^{w_+}dwe^{-3A(w)}
  \int_w^{w_+}dw'\frac{\dot{A}(w')^2}{\ddot{A}(w')+\dot{A}(w')^2}
  e^{3A(w')}\int_{w_-}^{w'}dw''e^{-3A(w'')} \nonumber\\
 & & +\frac{3e^{-3A_+}}{2\kappa_5^2\dot{\Psi}^{(0)}_+C_+}
  \left[1+e^{3A_+}(2\dot{A}_++{\alpha'}_+^{(0)}\dot{\Psi}^{(0)}_+)
   \int_{w_-}^{w_+}dwe^{-3A(w)}\right]^2
  -\frac{3e^{-3A_-}}{2\kappa_5^2\dot{\Psi}^{(0)}_-C_-}, \nonumber\\
 l_{T+}^2\times\Omega_+^{-2}\int_{w_-}^{w_+}dwe^{-3A(w)} & = & 
  \int_{w_-}^{w^+}dwe^{-3A(w)}
  \int_{w}^{w_+}dw'e^{3A(w')}\int_{w_-}^{w'}dw''e^{-3A(w'')}, 
  \nonumber\\
 \label{eqn:GN-lS-lT}
\end{eqnarray}
From these results, it is easy to show that
%
\begin{eqnarray}
 \left[ 1+l_{S+}^2\bar{q}_+^{(0)\mu\nu}k_{\mu}k_{\nu} + O(\mu^2)\right]
  \bar{f}_+ & = & 
  -16\pi G_{N+}\Omega_+^2\bar{\tau}_{+(LL)}, \nonumber\\
 \bar{q}_+^{(0)\mu\nu}k_{\mu}k_{\nu}
 \left[ 1-l_{T+}^2\bar{q}_+^{(0)\mu\nu}k_{\mu}k_{\nu} + O(\mu^2)\right]
 \bar{f}_{+(T)} & = & 16\pi G_{N+}\bar{\tau}_{+(T)},
 \label{eqn:result}
\end{eqnarray}
where $\bar{q}_+^{(0)\mu\nu}=\Omega_+^{-2}\eta^{\mu\nu}$ is the inverse of
the unperturbed physical metric, 
$\bar{q}_{+\mu\nu}^{(0)}=\Omega_+^2\eta_{\mu\nu}$, on $\Sigma_+$.

As promised, the length scale $l$ does not appear in the results
(\ref{eqn:result}). However, we have obtained two lengths $|l_{S+}|$ and
$l_{T+}$. It is evident that we cannot trust the $\mu$-expansion, or the
low energy expansion, at energies above
$\min(|l_{S\pm}^{-1}|,l_{T\pm}^{-1})$. Hence, we should choose
$l=\Omega_+^{-1}\max(|l_{S\pm}|,l_{T\pm})$.

The result (\ref{eqn:result}) is of the expected form 
(\ref{eqn:expectation}) with the expansion (\ref{eqn:expansion-C}) of
the functions $C_{(S,T)}$. We have determined first two coefficients of
the expansions:
%
\begin{eqnarray}
 C_{(S)}^{[0]} & = & -\frac{\Omega_+^{-2}}{16\pi G_{N+}}, \nonumber\\
 C_{(T)}^{[0]} & = & \frac{1}{16\pi G_{N+}}, \nonumber\\
 C_{(S)}^{[1]} & = & 
  \left(\frac{l_{S+}}{\Omega_+ l}\right)^2C_{(S)}^{[0]},   \nonumber\\
 C_{(T)}^{[1]} & = & 
  -\left(\frac{l_{T+}}{\Omega_+ l}\right)^2C_{(T)}^{[0]}. 
\end{eqnarray}


\section{Higher derivative gravity in four-dimensions} 
\label{sec:higher-derivative}

In this section, for the purpose of comparison, we review the linear
perturbations in $4$-dimensional higher derivative gravity. In
particular, we consider a theory whose action includes curvature-squared
terms: 
%
\begin{equation}
 I = \frac{1}{16\pi G_N}\int d^4x\sqrt{-g}
  \left[ R + \tilde{a}_1R^2 + \tilde{a}_2R^{\mu}_{\nu}R_{\mu}^{\nu} + 
   \tilde{a}_3R^{\mu\nu}_{\ \ \rho\sigma}
   R^{\rho\sigma}_{\ \ \mu\nu} \right],
  \label{eqn:4Daction1}
\end{equation}
where $\tilde{a}_1$, $\tilde{a}_2$ and $\tilde{a}_3$ are constants. 
The variation of this action plus matter action with respect to the
metric gives the following equation of motion. 
%
\begin{equation}
 G_{\mu\nu}+\tilde{a}_1E_{1\mu\nu}+\tilde{a}_2E_{2\mu\nu}
  +\tilde{a}_3E_{3\mu\nu} = 8\pi G_N T_{\mu\nu},
  \label{eqn:EOM1}
\end{equation}
where
%
\begin{eqnarray}
 E_{1\mu\nu} & = & -2\nabla_{\mu}\nabla_{\nu}R 
  + 2\nabla^2Rg_{\mu\nu} +2RR_{\mu\nu} - \frac{1}{2}R^2g_{\mu\nu}, 
  \nonumber\\
 E_{2\mu\nu} & = & -\nabla_{\mu}\nabla^{\rho}R_{\nu\rho}
  - \nabla_{\nu}\nabla^{\rho}R_{\mu\rho} + \nabla^2R_{\mu\nu}
  +\nabla^{\rho}\nabla^{\sigma}R_{\rho\sigma}g_{\mu\nu}
  +2R_{\mu\rho}R^{\rho}_{\nu} 
  - \frac{1}{2}R^{\rho}_{\sigma}R_{\rho}^{\sigma}g_{\mu\nu}, 
  \nonumber\\
 E_{3\mu\nu} & = & 2(\nabla^{\rho}\nabla^{\sigma} 
  + \nabla^{\sigma}\nabla^{\rho})R_{\mu\rho\nu\sigma}
  + 2R_{\mu\lambda\rho\sigma}R_{\nu}^{\ \lambda\rho\sigma}
  - \frac{1}{2}R^{\alpha\beta}_{\ \ \rho\sigma}
  R_{\ \ \alpha\beta}^{\rho\sigma}g_{\mu\nu}. 
\end{eqnarray}
It is known that these three tensors $E_{i\mu\nu}$ ($i=1,2,3$) are not
independent. Actually, it can be shown that
$E_{1\mu\nu}-4E_{2\mu\nu}+E_{3\mu\nu}=0$ in general. The easiest way to
see this relation is to note that the choice $(a_1,a_2,a_3)=(a,-4a,a)$
leads to a combination called Gauss-Bonnet term. Hence, the relation
$E_{1\mu\nu}-4E_{2\mu\nu}+E_{3\mu\nu}=0$ follows from the fact that the
Gauss-Bonnet term is topological. Because of the linear relation among
$E_{i\mu\nu}$ ($i=1,2,3$) it seems convenient to reparameterize the
action (\ref{eqn:4Daction1}). For later convenience we adopt the
following reparameterization by taking another combination
$3R^{\mu}_{\nu}R_{\mu}^{\nu}-R^2$. 
%
\begin{equation}
 I = \frac{1}{16\pi G_N}\int d^4x\sqrt{-g}
  \left[ R + a_1R^2 + a_2(3R^{\mu}_{\nu}R_{\mu}^{\nu}-R^2) + 
   a_3(R^{\mu\nu}_{\ \ \rho\sigma}R^{\rho\sigma}_{\ \ \mu\nu} 
   -4R^{\mu}_{\nu}R_{\mu}^{\nu}+R^2) \right],
  \label{eqn:4Daction2}
\end{equation}
where $a_1=\tilde{a}_1-\tilde{a}_2+\tilde{a}_3$,
$a_2=3\tilde{a}_2-4\tilde{a}_3$ and $a_3=\tilde{a}_3$. 
The equation of motion (\ref{eqn:EOM1}) becomes
%
\begin{equation}
 G_{\mu\nu}+a_1E_{1\mu\nu}+a_2(3E_{2\mu\nu}-E_{1\mu\nu}) 
  = 8\pi G_N T_{\mu\nu}, 
  \label{eqn:reparametrization}
\end{equation}
which is explicitly independent of $a_3$.

We consider general perturbations around the Minkowski
spacetime. Namely, we consider the metric of the form
%
\begin{equation}
 ds_4^2 = (\bar{q}^{(0)}_{\mu\nu}+\delta \bar{q}_{\mu\nu})
  dy^{\mu}dy^{\mu}, \nonumber\\
\end{equation}
where
%
\begin{equation}
 \bar{q}_{\mu\nu}^{(0)}  =  \Omega^2\eta_{\mu\nu}, 
\end{equation}
$\Omega$ is a non-zero constant, and 
%
\begin{equation}
 \delta \bar{q}_{\mu\nu} = 
   \bar{\sigma}_{(T)}T_{(T)\mu\nu} 
   + \bar{\sigma}_{(LT)}T_{(LT)\mu\nu}
   + \bar{\sigma}_{(LL)}T_{(LL)\mu\nu} 
   + \bar{\sigma}_{(Y)}T_{(Y)\mu\nu}.
\end{equation}
Here, the coefficients $\sigma_{(T)}$, $\sigma_{(LT)}$,
$\sigma_{(LL)}$, $\sigma_{(Y)}$ are constants. For the definition of the
harmonics $T_{(T,LT,LL,Y)\mu\nu}$, see appendix
\ref{app:harmonics}. Similarly to (\ref{eqn:fT-f}), we can construct
gauge-invariant variables $\bar{f}_{(T)}$ and $\bar{f}$. 
%
\begin{eqnarray}
 \bar{f}_{(T)} & = & \bar{\sigma}_{(T)}, \nonumber\\
 \bar{f} & = & \bar{\sigma}_{(Y)} 
  + \frac{1}{2}\eta^{\mu\nu}k_{\mu}k_{\nu}\bar{\sigma}_{(LL)}. 
\end{eqnarray}

As for the stress energy tensor $\bar{S}_{\mu\nu}$, we consider it
as a first order quantity, and expand it as follows. 
%
\begin{equation}
 \bar{S}_{\mu\nu} = 
  \bar{\tau}_{(T)}T_{(T)\mu\nu} + \bar{\tau}_{(LT)}T_{(LT)\mu\nu} 
   + \bar{\tau}_{(LL)}T_{(LL)\mu\nu} + \bar{\tau}_{(Y)}T_{(Y)\mu\nu}, 
\end{equation}
where coefficients $\bar{\tau}_{(T,LT,LL,Y)}$ are constants. In the
Minkowski background these coefficients are gauge-invariant by
themselves.

We can expand the equation of motion up to the first order 
in the perturbations and express it in terms of the above
gauge-invariant variables. The equation of motion is
%
\begin{eqnarray}
 2\bar{\tau}_{(Y)} & = & 
  3\eta^{\mu\nu}k_{\mu}k_{\nu}\bar{\tau}_{(LL)}, \nonumber\\
 \left[1 + 6a_1\bar{q}^{(0)\mu\nu}k_{\mu}k_{\nu}
\right]\bar{f} & = & -16\pi G_N\Omega^2\bar{\tau}_{(LL)}
 \label{eqn:HD-scalar}
\end{eqnarray}
for scalar perturbations, 
%
\begin{equation}
 \bar{\tau}_{(LT)} = 0
\end{equation}
for vector perturbations, and 
%
\begin{equation}
 \bar{q}^{(0)\mu\nu}k_{\mu}k_{\nu}
  \left[1-3a_2\bar{q}^{(0)\mu\nu}k_{\mu}k_{\nu}\right]
  \bar{f}_{(T)} 
  = 16\pi G_N\bar{\tau}_{(T)}
  \label{eqn:HD-tensor} 
\end{equation}
for tensor perturbations.

Therefore, the weak gravity equations (\ref{eqn:result}) (and the first
of (\ref{eqn:scalar-junction}))on our brane is indistinguishable from
the linearized gravitational equations (\ref{eqn:HD-scalar}) and
(\ref{eqn:HD-tensor}) in the higher derivative gravity, provided that
the following correspondence is understood. 
%
\begin{eqnarray}
 G_N & \leftrightarrow & G_{N+}, \nonumber\\
 6a_1 & \leftrightarrow & l_{S+}^2, \nonumber\\
 3a_2 & \leftrightarrow & l_{T_+}^2,
  \label{eqn:correspondence}
\end{eqnarray}
where $G_{N+}$, $l_{S+}^2$ and $l_{T+}^2$ are given by
(\ref{eqn:GN-lS-lT}).


\section{Physical implications}
\label{sec:implications}


The result of the previous sections up to the order $O(\mu)$ can be
summarized by the following $4$-dimensional effective gravitational
action on our brane.
%
\begin{equation}
 I = \frac{1}{16\pi G_{N\pm}}\int d^4y\sqrt{-\bar{q}_{\pm}}
  \left[ \bar{R}_{\pm} + \frac{1}{6}l_{S\pm}^2\bar{R}_{\pm}^2 
   + \frac{1}{3}l_{T\pm}^2(3\bar{R}^{\mu}_{\pm\nu}\bar{R}_{\pm\mu}^{\nu}
   -\bar{R}_{\pm}^2) + 
   a_{\pm}
   (\bar{R}^{\mu\nu}_{\pm\ \rho\sigma}\bar{R}^{\rho\sigma}_{\pm\ \mu\nu} 
   -4\bar{R}^{\mu}_{\pm\nu}\bar{R}_{\pm\mu}^{\nu}+\bar{R}^2_+) \right],
  \label{eqn:4Daction3}
\end{equation}
where $\bar{R}_{\pm}$, $\bar{R}^{\mu}_{\pm\nu}$ and 
$\bar{R}^{\mu\nu}_{\pm\ \rho\sigma}$ are the Ricci scalar, Ricci tensor 
and Riemann tensor of the minimally coupled physical metric
$\bar{q}_{\pm\mu\nu}$ on the brane $\Sigma_{\pm}$, and constants
$G_{N\pm}$, $l_{S\pm}^2$ and $l_{T\pm}^2$ are given by
(\ref{eqn:GN-lS-lT}) (and corresponding expressions for quantities with
the subscript ``$-$''). The expression with the undetermined coefficient
$a_{\pm}$ is the Gauss-Bonnet term and does not contribute to the
equations of motion at all. Hence $4$-dimensional Einstein gravity is 
restored only at distances much longer than
$\max(|l_{S\pm}|,l_{T\pm})$ or at energies much lower than
$\min(|l_{S\pm}^{-1}|,l_{T\pm}^{-1})$. Otherwise, $4$-dimensional
Einstein gravity is not valid on the brane $\Sigma_{\pm}$. What governs
weak gravity on the brane at higher energies is the higher derivative
gravity. Note that the energy scale
$\min(|l_{S\pm}^{-1}|,l_{T\pm}^{-1})$ can be lower 
than both $5$-dimensional and $4$-dimensional Planck energies.


The restoration of the higher derivative gravity on a brane was almost 
insensitive to the form of potentials of the scalar field and conformal
transformation to a frame in which matter on the branes is minimally
coupled to the metric. Only condition which we had to impose is that
there is a background solution with $4$-dimensional Poincare symmetry,
that $\dot{\Psi}_{\pm}^{(0)}C_{\pm}\ne 0$ and that 
$\dot{A}_{\pm}\ne 0$. The condition $\dot{\Psi}_{\pm}^{(0)}C_{\pm}\ne 0$
can be rewritten as follows~\cite{Mukohyama-Kofman}. 
%
\begin{equation}
 V^{'(0)}_{\pm}
  \left({U'_{\pm}}^{(0)}
     +\frac{\kappa_5^2}{3}V_{\pm}^{(0)}{V'_{\pm}}^{(0)}
                 -\frac{1}{4}{V'_{\pm}}^{(0)}{V''_{\pm}}^{(0)}\right)
  \ne 0.
  \label{eqn:assumption-potential}
\end{equation}
This condition is equivalent to the condition that the background bulk
scalar field is changing near the branes and that the
inter-brane distance (radion) is stabilized. To see this it is enough to
differentiate the background scalar-field matching condition
(the second of (\ref{eqn:background-junction})) with respect to
$w_{\pm}$. In fact, the radion stabilization can be stated as 
%
\begin{eqnarray}
 0 \neq e^{A_{\pm}}\frac{\partial}{\partial w_{\pm}}
  \left[\left.e^A\dot{\Psi}^{(0)}\right|_{w=w_{\pm}} 
  \pm\frac{1}{2}V'_{\pm}(\Psi^{(0)}_{\pm})\right]
   = {U'_{\pm}}^{(0)}
     +\frac{\kappa_5^2}{3}V_{\pm}^{(0)}{V'_{\pm}}^{(0)}
                 -\frac{1}{4}{V'_{\pm}}^{(0)}{V''_{\pm}}^{(0)}
\end{eqnarray}
since the following equation is always satisfied because of the
background equation. 
%
\begin{eqnarray}
 \frac{\partial}{\partial w_{\pm}}
 \left[ \left.e^A\dot{A}\right|_{w=w_{\pm}} 
  \pm\frac{1}{6}\kappa_5^2V_{\pm}(\Psi^{(0)}_{\pm})\right]=0.
\end{eqnarray}
Hence, actually the condition (\ref{eqn:assumption-potential}) is
equivalent to the condition that $V^{'(0)}_{\pm}\ne 0$ and that the
radion is stabilized. Since $V^{'(0)}_{\pm}$ is related to
$\dot{\Psi}^{(0)}$ by (\ref{eqn:background-junction}) on branes, finally
the condition (\ref{eqn:assumption-potential}) is equivalent to the
condition that the background bulk scalar field is changing near the
branes and that the radion is stabilized. The final condition
$\dot{A}_{\pm}\ne 0$ can be restated that the bulk geometry should be
warped near the branes.


A higher derivative theory with the action (\ref{eqn:4Daction3}) is
known to be unstable if the coefficient $l_{S\pm}^2$ is
negative~\cite{Muller-Schmidt}. One can see this instability in the
first equation of (\ref{eqn:result}) at the linearized level: if
$l_{S+}^2$ is negative then the equation has a tachyonic solution
($\bar{q}_+^{(0)\mu\nu}k^{\mu}k_{\nu}=|l_{S+}|^{-2}$) with vanishing
matter on the brane ($\bar{\tau}_{+(LL)}=0$). Since the coefficient
$l_{S\pm}^2$ depends on the background solution as shown in
(\ref{eqn:GN-lS-lT}), one can perhaps conclude that the stability
condition $l_{S\pm}^2\geq 0$ should be imposed as a constraint on models
of the brane-world. However, while we adopted the expansion in the
parameter $\mu$, the tachyonic solution corresponds to $\mu$ of order
unity. Hence, this solution is outside the domain of validity of the
expansion in $\mu$. In other words, terms of more than the second power
of $\mu$ can alter the stability/instability significantly. Further
investigation is necessary to understand the stability of the brane
world.

One could also derive another stability condition at the linearized
level from the second equation of (\ref{eqn:result}). The stability
condition at the linearized level would be $l_{T\pm}^2\leq 0$. 
Surprisingly, this condition is always violated since $l_{T\pm}^2$ is 
positive as shown in (\ref{eqn:GN-lS-lT}). Hence, the second equation
of (\ref{eqn:result}) has a tachyonic solution
($\bar{q}_+^{(0)\mu\nu}k^{\mu}k_{\nu}=l_{T+}^{-2}$) with vanishing 
matter on the brane ($\bar{\tau}_{+(T)}=0$). The appearance of this
tachyonic solution may be considered as the brane-world version of
Horowitz instability~\cite{Horowitz}: noting that Horowitz instability
is caused by the Weyl-squared term 
$C^{\mu\nu}_{\rho\sigma}C^{\rho\sigma}_{\mu\nu}$ in a semiclassical
effective action and that 
$C^{\mu\nu}_{\rho\sigma}C^{\rho\sigma}_{\mu\nu}= 
 (2/3)(3R^{\mu}_{\nu}R^{\nu}_{\mu}-R^2)
 +(\mbox{the Gauss-Bonnet term})$, 
one can expect that the term 
$(3\bar{R}^{\mu}_{\pm\nu}\bar{R}_{\pm\mu}^{\nu} -\bar{R}_{\pm}^2)$ in 
the action (\ref{eqn:4Daction3}) may cause an analogue of Horowitz
instability. However, the tachyonic solution in this case is also
outside the domain of validity of the expansion in $\mu$. Hence, terms
of more than the second power of $\mu$ can alter the instability
significantly. From this point of view we expect that the tachyonic
solution is just a spurious solution and does not indicate any
physically harmful instability of the brane world. Actually, it can be
shown that $\eta^{\mu\nu}k_{\mu}k_{\nu}\leq 0$ for any solutions of the
full equation (\ref{eqn:tensor-bulkeq}) insofar as the boundary
condition (\ref{eqn:bound-cond-tensor}) with $\bar{\tau}_{\pm(T)}=0$ is
satisfied~\cite{Kofman}~\footnote{For a similar statement without
branes, see ref.~\cite{DFGK}.}. The proof is easy: we can show that 
%
\begin{equation}
 \eta^{\mu\nu}k_{\mu}k_{\nu}\int_{w_-}^{w_+}dwe^AF_{(T)}^2
  = -\int_{w_-}^{w_+}dw
  e^{-3A}\left[(e^{2A}F_{(T)})^{\cdot}\right]^2 \leq 0.
\end{equation}
Therefore, we conclude that the brane-world version of Horowitz
instability is spurious and does not indicate any physically harmful
instability of the brane world.

Hence, we have to remove the spurious tachyonic solution for tensor
perturbations whenever we deal with the low energy equation
(\ref{eqn:result}) or the low energy effective action
(\ref{eqn:4Daction3}). One of the possible ways to remove spurious
solutions is the method of ``self-consistent reduction of order''. This
method has been used in many areas of physics including the radiation
reaction equation~\cite{Landau-Lifshitz}, the post-Newtonian equations
of motion in classical relativity~\cite{Damour}, higher derivative
gravity~\cite{Bel-Zia}, and the semiclassical
gravity~\cite{Parker-Simon} (see also~\cite{Flanagan-Wald}).


For some purposes it is useful to rewrite the gauge-invariant 
equations (\ref{eqn:result}) in a gauge-fixed form. We adopt a gauge
(the harmonic gauge) such that $\partial_{\mu}\bar{h}^{\mu}_{\nu}=0$, 
where $\bar{h}^{\mu}_{\nu}\equiv\delta\bar{q}^{\mu}_{\pm\nu} 
-\delta\bar{q}_{\pm}\delta^{\mu}_{\nu}/2$ and
$\delta\bar{q}_{\pm}=\delta\bar{q}^{\mu}_{\pm\mu}$. Hereafter, indices
are raised by $\bar{q}^{(0)\mu\nu}_{\pm}=\Omega_{\pm}^{-2}\eta^{\mu\nu}$
and lowered by $\bar{q}^{(0)}_{\pm\mu\nu}=\Omega_{\pm}^2\eta_{\mu\nu}$.
In this gauge the scalar-type gauge invariant variable $\bar{f}_{\pm}$
is reduced to $\bar{f}_{\pm}=(2/3)\bar{\sigma}_{\pm(Y)}$, and the first
equation of (\ref{eqn:result}) becomes 
%
\begin{equation}
 \bar{q}^{(0)\mu\nu}_{\pm}k_{\mu}k_{\nu}
  \left[ 1+l_{S\pm}^2\bar{q}_{\pm}^{(0)\mu\nu}k_{\mu}k_{\nu} 
   + O(\mu^2)\right]
  \bar{\sigma}_{\pm(Y)} = -16\pi G_N\bar{\tau}_{\pm(Y)}. 
\end{equation}
Hence, combining this equation with the second equation of
(\ref{eqn:result}) and the conservation equation (the first equation of
(\ref{eqn:scalar-junction})) we obtain 
%
\begin{equation}
 \Box\bar{h}^{\mu}_{\nu}+l_{T\pm}^2\Box^2\bar{h}^{\mu}_{\nu}
  +\frac{1}{3}(l_{T\pm}^2+l_{S\pm}^2)
  (\partial^{\mu}\partial_{\nu}\Box-\delta^{\mu}_{\nu}\Box^2)\bar{h}
  = -16\pi G_{N\pm}\bar{S}^{\mu}_{\pm\nu},
  \label{eqn:result-gauge-fixed}
\end{equation}
where $\bar{S}^{\mu}_{\pm\nu}$ is the surface energy momentum tensor of
matter on the brane $\Sigma_{\pm}$ in the minimally coupled physical
frame, $\Box=\partial^{\mu}\partial_{\mu}$, and
$\bar{h}=\bar{h}^{\mu}_{\mu}=-\delta\bar{q}_{\pm}$. Alternatively, the
equation (\ref{eqn:result-gauge-fixed}) can be derived from the action 
(\ref{eqn:4Daction3}) by using the well-known formula
$\bar{R}^{\mu}_{\pm\nu}= 
-\Box(\bar{h}^{\mu}_{\nu}-\bar{h}\delta^{\mu}_{\nu}/2)/2+O(\bar{h}^2)$
and (\ref{eqn:reparametrization}) with $a_1=l_{S\pm}^2/6$ and
$a_2=l_{T\pm}^2/3$. In the low energy limit the gauge-fixed equation
(\ref{eqn:result-gauge-fixed}), of course, reduces to the linearized
Einstein equation $\Box\bar{h}^{\mu}_{\nu}=-16\pi
G_{N\pm}\bar{S}^{\mu}_{\pm\nu}$. Again, we have to remove the spurious
tachyonic solution for tensor perturbations whenever we deal with the
low energy gauge-fixed equation (\ref{eqn:result-gauge-fixed}) by
eg. the method of ``self-consistent reduction of order''.


Given the high energy correction to $4$-dimensional Einstein gravity, it
is possible to derive corrections to the Newtonian potential. Actually,
it is known that the spherically symmetric, static solutions of the 
linearized field equations in theories with curvature-squared terms are
combinations of Newtonian and Yukawa potentials~\cite{Stelle}. Hence,
the inclusion of the higher derivative terms indeed changes the short
distance behavior of gravity, while it does not change the long distance
behavior. For this reason, the correction presented in this paper is 
relevant for laboratory experiment on the validity of the Newtonian
force. On the other hand, the long distance correction obtained by
Garriga and Tanaka~\cite{Garriga-Tanaka} is relevant for astronomical
tests although they considered a different model (the Randall-Sundrum
infinite bulk model without a scalar field). Calculation of long
distance corrections to the Newtonian potential in the model with a bulk
scalar field between two branes is a worth while task as a future
work. Analogy between calculations of long distance quantum 
corrections~\cite{Donoghue} and the brane world calculation of long
distance corrections is also an interesting subject.


Equipped with the result up to the order $O(\mu)$, we conjecture that in
the order $O(\mu^N)$, weak gravity on the brane should be still
indistinguishable from a higher derivative gravity whose action includes
up to $(N+1)$-th power of curvature tensors. The coefficients of higher
derivative terms can be in principle calculated by using the iterative
results in section \ref{sec:expansion}, as we have done in this paper up
to the order $O(\mu)$. Since the expansion in $\mu$ is in principle an
infinite series, gravity in the brane world becomes non-local at high
energies even at the linearized level. Actually, in this case the left
hand side of (\ref{eqn:result-gauge-fixed}) will have up to $2(N+1)$-th
derivatives ($N\to\infty$) and, thus, we need to specify the $0$-th to
$(2N+1)$-th time-derivatives ($N\to\infty$) of the field
$\bar{h}_{\mu\nu}$ on a spacelike $3$-surface in order to predict the
future evolution. In other words, we need to specify the whole (past)
history of $\bar{h}_{\mu\nu}$ to predict its future evolution, provided
that $\bar{h}_{\mu\nu}$ can be Taylor expanded with respect to the
time. This explains how the $4$-dimensional local description breaks
down at high energies. Of course, at low energies below
$\min(|l_{S\pm}|^{-1},l_{T\pm}^{-1})$ we expect that the non-local
behavior is suppressed.


It is known that if a gravitational action includes a metric tensor
(without derivatives) and its Ricci tensor only, then the system is
equivalent to another system described by Einstein gravity and
additional massive spin-$2$ and spin-$0$ fields~\cite{Maeda}. Evidently,
this observation is applicable to the effective action
(\ref{eqn:4Daction3}). Hence, we can expect appearance of effective
spin-$2$ and spin-$0$ fields on the branes. The appearance of these
fields is consistent with the observation that the Yukawa potential
appears in the correction to Newtonian potential by curvature-squared
terms as a result of exchanges of effective massive spin-$2$ and
spin-$0$ fields~\cite{Stelle}. In the brane world context, these
spin-$2$ and spin-$0$ fields should be composite fields due to
superpositions of bulk gravitational and scalar fields, respectively.


The appearance of the composite spin-$0$ field may suggest a possibility
of brane-world inflation without an inflaton on the brane. Actually, in
$4$-dimensional higher derivative gravitational theories it is known
that inflation can be driven by higher derivative terms. Such an
inflationary model is called Starobinsky
model~\cite{Starobinsky,Vilenkin,KMP}. Hence, there is a possibility 
that the brane-world version of Starobinsky model may be caused by the
higher derivative terms found in this paper. In this case, the composite
spin-$0$ field is the brane-world version of the so called scalaron. 
This possibility of brane-world inflation due to higher derivative
terms seems to be related to an interesting brane inflation scenario
proposed by Himemoto and Sasaki~\cite{Himemoto-Sasaki}. It seems
interesting to explore relations between their model and the brane-world
version of Starobinsky model. It may be sufficient to suppose that brane
inflation is driven by the composite spin-$0$ field, or the brane-world
version of the scalaron. It is expected that fluctuations of the
spin-$2$ composite field as well as the spin-$0$ field (scalaron) play
important roles in the generation of cosmological perturbations in this
case.


Although the composite spin-$2$ and spin-$0$ fields are generally
expected to appear on the brane, it is not the end of the story. As
far as the author knows, there is no known equivalence between a higher
derivative gravity theory and Einstein theory with additional fields if
the higher derivative terms in the action cannot be expressed in terms
of Ricci tensor only. In other words, if the higher derivative terms
depend on the Weyl tensor explicitly then the composite spin-$2$ and
spin-$0$ fields are not enough to describe the whole system. Actually,
the non-locality discussed above seems to require that the appearance of
these composite fields is not the whole story since these fields have
$4$-dimensional local actions.


The expected non-locality should be due to gravitational and scalar
waves in the bulk. Hence, the infinite series of higher derivative terms
is one description of the non-locality pointed out in
ref.~\cite{Mukohyama2000b} in the context of brane world 
cosmology~\cite{Cosmology} (see also \cite{Alvarez-Mazzitelli}). Another
description was given as an integro-differential equation in
ref.~\cite{Mukohyama2001a}, where a complete set of four equations
governing scalar-type cosmological perturbations was derived by using
the doubly gauge invariant formalism developed in
refs.~\cite{Mukohyama2000b,Mukohyama2000c}. One of the four equations is
an integro-differential equation, which describes non-local effects due
to gravitational waves propagating in the bulk. Further investigation of
the relation between the two different descriptions may be an
interesting future subject.


\section{Summary}
\label{sec:summary}


In summary we have extended the analysis of linearized gravity in brane
world models with a bulk scalar field between two branes to higher
energies by investigating the next relevant order in the expansion in
the parameter $\mu=l^2\eta^{\mu\nu}k_{\mu}k_{\nu}$, where $l$
is the characteristic length scale of the model given by
$l=\Omega_+^{-1}\max(|l_{S\pm}|,l_{T\pm})$. Since the $4$-dimensional
physical energy scale $m_+$ on our brane $\Sigma_+$ is given by 
$m_+^2=-\Omega_+^{-2}l^{-2}\mu$, the expansion in $\mu$ is nothing but
the low energy expansion. For the investigation we used the formalism
developed in ref.~\cite{Mukohyama-Kofman}, in which all quantities and
equations including the surface energy momentum tensor and the junction
condition are completely Fourier transformed with respect to the
$4$-dimensional coordinates so that the problem is essentially reduced
to a set of $1$-dimensional problems.

We compared the result with the so called higher derivative gravity. It
has been shown that in the order $O(\mu)$, gravity on the brane is
indistinguishable from the higher derivative gravity whose action
includes the Einstein term and curvature-squared terms, provided that
the inter-brane distance (radion) is stabilized, that the background
scalar field is changing near the branes and that the background bulk
geometry near the branes is warped. This result holds for a general
conformal transformation to a frame in which matter on the branes is
minimally coupled to the metric. The obtained indistinguishability
between brane gravity and higher derivative gravity agrees with
ref.~\cite{Tekin} in which the non-relativistic limit of the theories
with non-warped extra dimension was considered. In the present paper we
considered weak gravity (including the non-relativistic and relativistic
limits) in the brane world model with a scalar field in a warped bulk 
geometry. Newton's constant and coefficients of curvature-squared terms
except for the Gauss-Bonnet topological term have been determined as in
(\ref{eqn:correspondence}). (The coefficient of the $4$-dimensional
Gauss-Bonnet term cannot and does not need to be determined as far as
classical dynamics is concerned.) The result is summarized by the
$4$-dimensional effective action (\ref{eqn:4Daction3}). In other words,
we have provided the brane-world realization of the so called
$R^2$-model.

Equipped with the result up to the order $O(\mu)$, we conjectured that 
in the order $O(\mu^N)$, weak gravity on the brane is still
indistinguishable from a higher derivative gravity whose action includes
up to $(N+1)$-th power of curvature tensors. We discussed the appearance
of composite spin-$2$ and spin-$0$ fields in addition to the graviton on
the brane and pointed out a possibility that the spin-$0$ field may play
the role of an effective inflaton to drive brane-world inflation. We
also showed that the brane-world version of Horowitz instability is
spurious. Finally, we conjectured that the sequence of higher derivative
terms is an infinite series and, thus, indicates non-locality in the
brane world scenarios.

\begin{acknowledgments}
 The author would like to thank Lev Kofman, Takahiro Tanaka and
 Alexander Vilenkin for useful discussions and valuable comments. He
 would be grateful to Werner Israel for continuing encouragement and
 careful reading of the manuscript. This work is supported by JSPS
 Postdoctoral Fellowship for Research Abroad. 
\end{acknowledgments}


\appendix


\section{Harmonics in Minkowski spacetime}
\label{app:harmonics}

In this appendix we give definitions of scalar, vector and tensor
harmonics in an $n$-dimensional Minkowski spacetime. Throughout this 
appendix, $n$-dimensional coordinates are $x^{\mu}$
($\mu=0,1,\cdots,n-1$), $\eta_{\mu\nu}$ is the Minkowski metric, and all
indices are raised and lowered by the Minkowski metric and its inverse
$\eta^{\mu\nu}$.

\subsection{Scalar harmonics}

The scalar harmonics are given by 
%
\begin{equation}
 Y = exp(-ik_{\rho}x^{\rho}),
\end{equation}
by which any function $f$ can be expanded as 
%
\begin{equation}
 f = \int dk\ c Y,
\end{equation}
where $c$ is a constant depending on $k$. Hereafter, $k$ and $dk$ are
abbreviations of $\{k^{\mu}\}$ ($\mu=0,1,\cdots,n-1$) and 
$\prod_{\mu=0}^{n-1}dk^{\mu}$, respectively. We omit $k$ in most cases. 

\subsection{Vector harmonics}

In general, any vector field $v_{\mu}$ can be decomposed as
%
\begin{equation}
 v_{\mu}=v_{(T)\mu}+\partial_{\mu} f ,
\end{equation}
where $f$ is a function and $v_{(T)\mu}$ is a transverse vector field:
%
\begin{equation}
 \partial^{\mu}v_{(T)\mu}=0 .
\end{equation}

Thus, the vector field $v_{\mu}$ can be expanded by using the scalar 
harmonics $Y$ and transverse vector harmonics $V_{(T)\mu}$ as 
%
\begin{equation}
 v_{\mu} = \int dk
  \left[c_{(T)}V_{(T)\mu}+c_{(L)}\partial_{\mu} Y\right].
        \label{eqn:dY+V}
\end{equation}
Here, $c_{(T)}$ and $c_{(L)}$ are constants depending on $k$, and the
transverse vector harmonics $V_{(T)\mu}$ are given by 
%
\begin{equation}
 V_{(T)\mu} = u_{\mu}\exp(-ik_{\rho}x^{\rho}),
\end{equation}
where the constant vector $u_{\mu}$ satisfies the following condition. 
%
\begin{equation}
 k^{\mu}u_{\mu}=0
\end{equation}
for $k^{\mu}k_{\mu}\ne 0$, and 
%
\begin{eqnarray}
 k^{\mu}u_{\mu} & = & 0, \nonumber\\
 \tau^{\mu}u_{\mu} & = & 0
  \label{eqn:u-for-k^2=0}
\end{eqnarray}
for non-vanishing $k_{\mu}$ satisfying $k^{\mu}k_{\mu}=0$, where
$\tau^{\mu}$ is an arbitrary constant timelike vector. 
For $k_{\mu}=0$,
the constant vector $u^{\mu}$ does not need to satisfy any of the above
conditions. For the special case $k^{\mu}k_{\mu}=0$, the second
condition in (\ref{eqn:u-for-k^2=0}) can be imposed by redefinition of
$c_{(L)}$. Actually this condition is necessary to eliminate
redundancy. Note that the number of independent vectors satisfying the
above condition is $n-1$ for $k^{\mu}k_{\mu}\ne 0$ and $n-2$ for
$k^{\mu}k_{\mu}=0$ and that these numbers are equal to the numbers of
physical degrees of freedom for massive and massless spin-$1$ fields in
$n$-dimensions, respectively.

Because of the expansion (\ref{eqn:dY+V}), it is convenient to define
longitudinal vector harmonics $V_{(L)\mu}$ by
%
\begin{equation}
 V_{(L)\mu} \equiv \partial_{\mu} Y = -ik_{\mu}Y. 
\end{equation}

\subsection{Tensor harmonics}

In general, a symmetric second-rank tensor field $t_{\mu\nu}$ can be
decomposed as
%
\begin{equation}
 t_{\mu\nu}=t_{(T)\mu\nu} + \partial_{\mu}v_{\nu}+\partial_{\nu}v_{\mu} 
  + f\eta_{\mu\nu},
\end{equation}
where $f$ is a function, $v_{\mu}$ is a vector field and $t_{(T)\mu\nu}$
is a transverse traceless symmetric tensor field:
%
\begin{eqnarray}
 t_{(T)\mu}^\mu & = & 0,\nonumber\\
 \partial^{\mu} t_{(T)\mu\nu} & = &0.
        \label{eqn:trasverse-traceless}
\end{eqnarray}

Thus, the tensor field $t_{\mu\nu}$ can be expanded by using the scalar
harmonics $Y$, the vector harmonics $V_{(T)}$ and $V_{(L)}$, and
transverse traceless tensor harmonics $T_{(T)}$ as 
%
\begin{eqnarray}
 t_{\mu\nu} & = & \int dk\left[
        c_{(T)}T_{(T)\mu\nu}+c_{(LT)}
        (\partial_{\mu}V_{(T)\nu}+\partial_{\nu}V_{(T)\mu})\right.
        \nonumber\\
 & &    \left.
         + c_{(LL)}(\partial_{\mu}V_{(L)\nu}+\partial_{\nu}V_{(L)\mu})
        + \tilde{c}_{(Y)}Y\eta_{\mu\nu}\right].
        \label{eqn:dV+T}
\end{eqnarray}
Here, $c_{(T)}$, $c_{(LT)}$, $c_{(LL)}$, and $\tilde{c}_{(Y)}$ are
constants depending on $k$, and the transverse traceless tensor
harmonics $T_{(T)}$ are given by 
%
\begin{equation}
 T_{(T)\mu\nu} = s_{\mu\nu}\exp(-ik_{\rho}x^{\rho}),
\end{equation}
where the constant symmetric second-rank tensor $s_{\mu\nu}$ satisfies
the following condition. 
%
\begin{eqnarray}
 k^{\mu}s_{\mu\nu} & = & 0, \nonumber\\
 s^{\mu}_{\mu} & = & 0
\end{eqnarray}
for $k^{\mu}k_{\mu}\ne 0$, and 
%
\begin{eqnarray}
 k^{\mu}s_{\mu\nu} & = & 0, \nonumber\\
 s^{\mu}_{\mu} & = & 0, \nonumber\\
 \tau^{\mu}s_{\mu\nu} & = & 0
  \label{eqn:s-for-k^2=0}
\end{eqnarray}

for non-vanishing $k_{\mu}$ satisfying $k^{\mu}k_{\mu}=0$, where
$\tau^{\mu}$ is an arbitrary constant timelike vector. 
For $k_{\mu}=0$, the constant tensor $s_{\mu\nu}$ does not need to
satisfy any of the above conditions. For the special case
$k^{\mu}k_{\mu}=0$, the last condition in (\ref{eqn:s-for-k^2=0}) can be
imposed by redefinition of $c_{(LT)}$, $c_{(LL)}$ and
$\tilde{c}_{(Y)}$. Actually this condition is 
necessary to eliminate redundancy. Note that the number of independent
symmetric second-rank tensors satisfying the above conditions is
$(n+1)(n-2)/2$ for $k^{\mu}k_{\mu}\ne 0$ and $n(n-3)/2$ for 
$k^{\mu}k_{\mu}=0$ and that these numbers are equal to numbers of
physical degrees of freedom for massive and massless spin-$2$ fields in
$n$-dimensions, respectively.

Because of the expansion (\ref{eqn:dV+T}), it is convenient to define
tensor harmonics $T_{(LT)}$, $T_{(LL)}$, and $T_{(Y)}$ by 
%
\begin{eqnarray}
 T_{(LT)\mu\nu} & \equiv & \partial_{\mu}V_{(T)\nu}
  +\partial_{\nu}V_{(T)\mu}, \nonumber\\
 & = & -i(u_{\mu}k_{\nu}+u_{\nu}k_{\mu})Y, \nonumber\\
 T_{(LL)\mu\nu} & \equiv & \partial_{\mu}V_{(L)\nu}
  +\partial_{\nu}V_{(L)\mu}
        -\frac{2}{n}\eta_{\mu\nu}\partial^{\rho}V_{(L)\rho} \nonumber\\
 & = & \left(-2k_{\mu}k_{\nu}+\frac{2}{n}k^{\rho}k_{\rho}
        \eta_{\mu\nu}\right)Y,  \nonumber\\ 
 T_{(Y)\mu\nu} & \equiv & \eta_{\mu\nu}Y. 
\end{eqnarray}



\begin{references}
 \bibitem{Polchinski}
 J. Polchinski, {\it String Theory I\& II} (Cambridge University Press, 
 1998).
 \bibitem{ADD}
 N.~Arkani-Hamed, S.~Dimopoulos and G.~Dvali, Phys. Rev. {\bf B429}, 263
 (1998); Phys. Rev.{\bf D59}, 086004 (1999). 
 \bibitem{AADD}
 L.~Antoniadis, N. Arkani-Hamed, S.~Dimopoulos and G.~Dvali,
 Phys. Lett. {\bf B436}, 257 (1998). 
 \bibitem{RS1}
 L.~Randall and R.~Sundrum, Phys. Rev. Lett. {\bf 83}, 3370 (1999).
 \bibitem{RS2}
 L.~Randall and R.~Sundrum, Phys. Rev. Lett. {\bf 83}, 4690 (1999).
 \bibitem{earlier-works}
 K.~Akama, in Lect. Notes Phys. {\bf 176}, Gauge Theory and
 Gravitation, Proceedings, Nara, 1982, edited by K.~Kikkawa,
 N.~Nakanishi and H.~Nariai (Springer-Verlag,1983) [hep-th/0001113]; 
 V.~A.~Rubakov and M.~E.~Shaposhnikov, Phys. Lett. {\bf 152B}, 136
 (1983); 
 M.~Visser, Phys. Lett. {\bf B159}, 22 (1985);
 \bibitem{Horava}
 P.~Horava and E.~Witten, Nucl. Phys. {\bf 460}, 506 (1996).
 \bibitem{Lukas}
 A.~Lukas, B.~A.~Ovrut, K.~S.~Stelle and D.~Waldram, Nucl.Phys. 
 {\bf B552}, 246 (1999).
 \bibitem{GW}
 W.~D.~Goldberger and M.~B.~Wise, Phys. Rev. Lett. {\bf 83}, 4922
 (1999). 
 \bibitem{DFGK}
 O.~DeWolfe, D.~Z.~Freedman, S.~S.~Gubser and A.~Karch, Phys. Rev. 
 {\bf D62}, 046008 (2000). 
 \bibitem{Garriga-Tanaka}
 J.~Garriga and T.~Tanaka, Phys. Rev. Lett. {\bf 84}, 2778 (2000).
 \bibitem{SSM}
 M.~Sasaki, T.~Shiromizu and K.~Maeda, Phys. Rev. {\bf D62}, 024008 
 (2000).
 \bibitem{GKR}
 S.~B.~Giddings, E.~Katz, and L.~Randall, J. High Energy Phys. {\bf 03},
 023 (2000).
 \bibitem{Tanaka-Montes}
 T.~Tanaka and X.~Montes, Nucl. Phys. {\bf B582}, 259 (2000). 
 \bibitem{Kudoh-Tanaka}
 H.~Kudoh and T.~Tanaka, Phys. Rev. {\bf D64}, 084022 (2001);
 hep-th/0112013. 
 \bibitem{Mukohyama-Kofman}
 S.~Mukohyama and L.~Kofman, hep-th/0112115.
 \bibitem{Mukohyama2001b}
 S.~Mukohyama, gr-qc/0108048, to appear in Phys. Rev. {\bf D}. 
 \bibitem{Stelle}
 K.~S.~Stelle, Gen. Rel. and Grav. {\bf 9}, 353 (1978). 
 \bibitem{Donoghue}
 J.~F.~Donoghue, Phys. Rev. Lett. {\bf 72}, 2996 (1994). 
 \bibitem{Maeda}
 K.~Maeda, Phys. Rev. {\bf D39}, 3159 (1989); 
 J.~Koga and K.~Maeda, Phys.Rev. {\bf D58}, 064020 (1998). 
 \bibitem{Himemoto-Sasaki}
 Y.~Himemoto and M.~Sasaki, Phys. Rev. {\bf D63}, 044015 (2001);
 N.~Sago, Y.~Himemoto and Misao Sasaki, gr-qc/0104033;
 Y.~Himemoto, T.~Tanaka and M.~Sasaki, gr-qc/0112027. 
 \bibitem{Starobinsky}
 A.~A.~Starobinsky, Phys. Lett. {\bf B91}, 99 (1980). 
 \bibitem{Vilenkin}
 A.~Vilenkin, Phys. Rev. {\bf D32}, 2511 (1985). 
 \bibitem{KMP}
 L.~Kofman, V.~Mukhanov and D.~Pogosyan, Sov. Phys. JETP {\bf 66}, 433
 (1987). 
 \bibitem{Muller-Schmidt}
 V.~M\"uller and H.-J.~Schmidt, Gen. Rel. and Grav. {\bf 17}, 769
 (1985); H.-J.~Schmidt, Phys. Rev. {\bf D50}, 5452 (1994).
 \bibitem{Horowitz}
 G.~T.~Horowitz, Phys. Rev. {\bf D21}, 1445 (1980). 
 \bibitem{Kofman}
 The author thanks Lev Kofman for pointing this out. 
 \bibitem{Landau-Lifshitz}
 For example, L.~D.~Landau and E.~M.~Lifshitz, {\it The classical Theory
 of Fields} (Pergamon, Oxford, 1962).
 \bibitem{Damour}
 For example, T.~Damour, in {\it Gravitational Radiation}, edited by
 N.~Deruelle and T.~Piran (North-Holland, Amsterdam, 1983), p.214.
 \bibitem{Bel-Zia}
 L.~Bel and H.~S.~Zia, Phys. Rev. {\bf D32}, 3128 (1985). 
 \bibitem{Parker-Simon}
 L.~Parker and J.~Z.~Simon, Phys. Rev. {\bf D47}, 1339 (1993). 
 \bibitem{Flanagan-Wald}
 \'E.~\'E.~Flanagan and R.~M.~Wald, Phys. Rev. {\bf D54}, 6233 (1996). 
 \bibitem{Mukohyama2000b}
 S.~Mukohyama, Phys. Rev. {\bf D62}, 084015 (2000).
 \bibitem{Cosmology}
 J.~M.~Cline, C.~Grojean and G.~Servant, Phys. Rev. Lett. 
 {\bf 83} 4245 (1999); 
 E.~E.~Flanagan, S.~H.~H.~Tye, I.~Wasserman, Phys. Rev. {\bf D62},
 044039 (2000);
 P.~Bin\'{e}truy, C.~Deffayet, U.~Ellwanger and D.~Langlois,
 Phys. Lett. {\bf B477}, 285 (2000); 
 S.~Mukohyama, Phys. Lett. {\bf B473}, 241 (2000); 
 P.~Kraus, J. High Energy Phys. {\bf 9912}, 011 (1999); 
 D.~Ida, JHEP {\bf 0009}, 014 (2000);
 S.~Mukohyama, T.~Shiromizu and K.~Maeda, Phys. Rev. {\bf D62}, 024028
 (2000), Erratum-ibid. {\bf D63}, 029901 (2001). 
 \bibitem{Alvarez-Mazzitelli}
 E.~Alvarez and F.~D.~Mazzitelli, Phys. Lett. {\bf B 505}, 236 (2001). 
 \bibitem{Mukohyama2001a}
 S.~Mukohyama, Phys. Rev. {\bf D64}, 064006 (2000).
 \bibitem{Mukohyama2000c}
 S.~Mukohyama, Class. Quantum Grav. {\bf 17}, 4777 (2000).
 \bibitem{Tekin}
 B.~Tekin, hep-th/0106134.
\end{references}
\end{document}